\documentclass[aps, prd, onecolumn, eqsecnum, amsmath, nofootinbib, preprintnumbers]{revtex4}%

\usepackage{color,graphicx,float,subfigure}
\usepackage{amsfonts,amssymb,theorem,mathrsfs,times}
\usepackage{bm}
\usepackage{mathtools}
\usepackage{amsfonts,amssymb,theorem,mathrsfs}
\usepackage{dsfont}
\usepackage{setspace}
\usepackage{cases}
\usepackage{listings}
\usepackage{tikz}
\usepackage{subfigure}
\usepackage{amsmath}
\usepackage{CJKutf8}
\usepackage{dutchcal}

\usepackage[colorlinks,linkcolor=blue,anchorcolor=blue,citecolor=blue]{hyperref}   
\usepackage{ulem}   

\newcommand{\checked}[1]{\textcolor{brown}{\bf{#1}}}

{\theorembodyfont{\upshape}
	}
{\theorembodyfont{\upshape}
	}
{\theorembodyfont{\upshape}
	}
{\theorembodyfont{\upshape}
	}
{\theorembodyfont{\upshape}
	}
{\theorembodyfont{\upshape}
	}
\addtocounter{MaxMatrixCols}{10}
\newcommand{\dalm}{\kern1pt\vbox{\hrule height 0.9pt\hbox{\vrule width
			0.9pt\hskip 2.5pt\vbox{\vskip 5.5pt}\hskip 3pt\vrule width
			0.3pt}\hrule height 0.3pt}\kern1pt}
\begin{document}

\preprint{\hfill {\small {ICTS-USTC/PCFT-23-22}}}
\title{Quasi normal modes of tensor perturbation in Kaluza-Klein black hole for Einstein-Gauss-Bonnet gravity}
	
%
	
\author{ Li-Ming Cao$^{a\, ,b}$\footnote{e-mail
			address: caolm@ustc.edu.cn}} 
		
\author{ Liang-Bi Wu$^b$\footnote{e-mail
			address: liangbi@mail.ustc.edu.cn}}
   
\author{ Yaqi Zhao$^c,^d$\footnote{e-mail
		address: zxmyg86400@mail.ustc.edu.cn}}
  
\author{ Yu-Sen Zhou$^b$\footnote{e-mail
		address: zhou\_ys@mail.ustc.edu.cn}}	
	
\affiliation{$^a$Peng Huanwu Center for Fundamental Theory, Hefei, Anhui 230026, China}
	
\affiliation{${}^b$Interdisciplinary Center for Theoretical Study and Department of Modern Physics,\\
	University of Science and Technology of China, Hefei, Anhui 230026,
	China}
 
\affiliation{${}^c$Deep Space Exploration Laboratory/School of Physical Science,
University of Science and Technology of China,
96 Jinzhai Road, Hefei, Anhui 230026, China}
	
\affiliation{${}^d$CAS Key Laboratory for Researches in Galaxies and Cosmology/Department of Astronomy, School of Astronomy and Space Science, University of Science and Technology of China, 96 Jinzhai Road, Hefei, Anhui 230026, China}	
	
\date{\today}
	
\begin{abstract}
 In Einstein–Gauss-Bonnet gravity, we study the quasi-normal modes (QNMs) of the tensor perturbation for the so-called Maeda-Dadhich  black hole which locally has a topology $\mathcal{M}^n \simeq M^4 \times \mathcal{K}^{n-4}$. Our discussion is based on the tensor perturbation equation derived in~\cite{Cao:2021sty}, where the Kodama-Ishibashi gauge invariant formalism for Einstein gravity theory has been generalized to the  Einstein–Gauss-Bonnet gravity theory. With the help of characteristic tensors for the constant curvature space $\mathcal{K}^{n-4}$, we investigate the effect of extra dimensions and obtain the scalar equation in four dimensional spacetime, which is quite different from the Klein-Gordon equation. Using the asymptotic iteration method and the numerical integration method with the Kumaresan-Tufts frequency extraction method, we numerically calculate the QNM frequencies. In our setups, characteristic frequencies depend on six distinct factors. They are the spacetime dimension $n$, the Gauss-Bonnet coupling constant $\alpha$, the black hole mass parameter $\mu$, the black hole charge parameter $q$, and two ``quantum numbers" $l$, $\gamma$. Without loss of generality, the impact of each parameter on the characteristic frequencies is investigated while fixing other five parameters. Interestingly, the dimension of compactification part has no significant impact on the lifetime of QNMs.
\end{abstract}

\maketitle
	
\section{Introduction}
\label{sec: introduction}
Modified gravity theories can help us understand the limitations of Einstein gravity theory, address challenges with existing models, and contribute significantly to our understanding of the Universe. Lovelock theories beyond four dimensions are the most general diffeomorphism covariant modified gravity theories only involving a metric tensor with second order equations of motion~\cite{Lovelock:1971yv}. In four dimensions and generic values of the coupling constants, the theory reduces to the General Relativity with a cosmological constant~\cite{Lovelock:1972vz}, where the corresponding equations of motion are the Einstein equations. Einstein-Gauss-Bonnet (EGB) gravity is the lowest Lovelock theory, whose Lagrangian contains only the linear and quadratic terms of spacetime curvature. String theory predicts quantum corrections to General Relativity, with the Gauss-Bonnet term which is the first and the dominating correction among others. EGB gravity is the simplest model for illustrating the distinctions between general Lovelock gravity theory and Einstein gravity theory in higher dimensions.

One of the best ways to understand the modified gravity theory is to investigate the black hole solutions within it. However, in modified gravity theories, analytical solutions of black holes are quite rare. Although numerical solutions are practical, it is necessary to have a certain foundation in numerical calculation theory in order to determine the accuracy of the results. Fortunately, several analytical black hole solutions in EGB gravity have been found. Black holes in high dimensional spacetime have garnered significant interest for two primary reasons: they arise naturally in the context of string theory and are also present in extra-dimensional brane-world scenarios~\cite{Randall:1999ee}. A class of static vacuum solutions in the EGB gravity was first obtained by Boulware, Deser, and Wheeler~\cite{Boulware:1985wk, Wheeler:1985nh}. This work was later extended to include a cosmological constant by Cai~\cite{Cai:2001dz}. The topology of these solutions is locally $\mathcal{M}^n \simeq M^2 \times \mathcal{N}^{n-2}$. The Maeda-Dadhich black hole, a kind of Kaluza-Klein (KK) black hole, is also an exact vacuum solution of EGB gravity with a cosmological constant which bears a specific relation to the Gauss-Bonnet coupling constant~\cite{Maeda:2006iw}. This spacetime is locally the product of a usual $4$-dimensional manifold with a $(n-4)$-dimensional space of constant negative curvature, i.e., its topology is locally $\mathcal{M}^n \simeq M^4 \times \mathcal{K}^{n-4}$. Here, $M^2$, $M^4$ and $\mathcal{N}^2$ are general (pseudo-)Riemannian manifold, while $\mathcal{K}^{n-4}$ is a non-compact Riemannian manifold with sectional curvature $K=-1$. Another remarkable feature of this solution is that the Gauss-Bonnet term acts like a Maxwell source for large $r$ while at the other end it regularizes the metric and weakens the central singularity~\cite{Maeda:2006iw}. It can serve as an excellent model for testing extra dimensions and could potentially be a candidate for a realistic black hole. Our present work will concentrate on this black hole solution. Based on the same ideas, a class of black hole solutions has been obtained in $n$-dimensional Lovelock gravity theory~\cite{Cai:2009de}. The topology of these solutions is locally $\mathcal{M}^n\simeq M^m\times\mathcal{K}^{n-m}$, where $\mathcal{K}^{n-m}$ is a space of negative constant curvature. 

Perturbing black holes provides valuable insights into their properties, but gauge dependence can be an issue. To address this, one approach is to use physically preferred gauges, while another one is to use gauge-invariant variables such as the Kodama-Ishibashi gauge invariant variables~\cite{Ishibashi:2011ws}. These variables allow for the derivation of master equations with tensor, vector, and scalar components~\cite{Ishibashi:2011ws,Dotti:2004sh,Cai:2013cja,Dotti:2005sq}. Using the Kodama-Ishibashi gauge invariant variables, a generalized master equation for tensor-type perturbations has been derived in EGB gravity~\cite{Cao:2021sty}.

When a black hole undergoes perturbations, it experiences damping oscillations superimposed by characteristic modes. The incoming boundary condition at the horizon and the outgoing boundary condition at spatial infinity result in a dissipative system with characteristic modes referred to as quasi-normal modes (QNMs). Their frequencies $\omega$, denoted as quasi-normal frequencies, are the group of discrete complex eigenvalues of perturbation equations of the black hole solution, a set of homogeneous second order differential equations, and can therefore reveal useful information about the corresponding spacetime geometry.

QNMs have been a subject of interest for several decades, ever since their initial proposal by Regge and Wheeler in their analysis of the stability of Schwarzschild black holes~\cite{Regge:1957td}. These modes are significant from both theoretical and observational standpoints. From the observation aspect, binary black hole mergers are a major source of gravitational waves. The waves emitted during the ringdown stage can be expressed as a superposition of quasi-normal modes of a perturbed Kerr black hole~\cite{Nollert:1999ji,Konoplya:2011qq,Isi:2021iql}. Some gravitational wave (GW) events have been detected by LIGO which are caused by the merger of binary BHs, where QNMs are also present~\cite{ligoscientific:2016aoc, ligoscientific:2016sjg, ligoscientific:2017bnn, ligoscientific:2017vwq}. These results show that the gravitational wave signal is consistent with Einstein gravity theory. However, the currently available observational precision of the binary BHs evolution prediction simulated by the LIGO and VIRGO collaborations may meet the requirements for testing modified gravity theories~\cite{ligoscientific:2017bnn}. That is to say, some non-negligible uncertain parameters within the BHs range indicate that the window for alternative gravity theories has been opened~\cite{Konoplya:2016pmh}. With the development of observational technology, it is now possible to use QNMs measurements from gravitational wave observations to test General Relativity, examine the validity of the ``no-hair" theorem~\cite{Dreyer:2003bv,Berti:2005ys,Isi:2019aib}, and constrain modified gravitational theories. Therefore, modified gravity theories containing higher derivative terms and extra dimensions may be truly confirmed through corresponding GW observations. As such, the analysis of QNMs has become an important topic in gravitational wave research.

From the theoretical perspective, QNMs are a topic of significant interest. Research into QNMs can provide insight into potential violations of strong cosmic censorship~\cite{Cardoso:2017soq}. Additionally, QNMs might provide one with some inspirations about the quantization of black hole area~\cite{Hod:1998vk,Dreyer:2002vy,Kunstatter:2002pj}. Furthermore, it is anticipated that the signatures of extra dimensions may be discerned from the QNMs of black holes. For instance, a recent study investigated the numerical evolution of massive Kaluza-Klein modes of a scalar field in a thick brane~\cite{Tan:2022uex}. This study found that there are scalar KK resonant particles with long lifespans on the brane, suggesting that these resonances could potentially serve as candidates for dark matter. Another study examined the quasi-normal modes of a thick brane in order to detect sounds from extra dimensions~\cite{Tan:2022vfe}. In addition, QNMs in EGB theory have also received a lot of attention~\cite{Moura:2021eln,Moura:2021nuh,Moura:2022gqm}. As for the Maeda-Dadhich black hole, Alexeyev and Petrov analyzed the stability of this black hole by using the Chandrasekhar frame~\cite{Alexeyev:2015mta}. However, their study was limited to perturbations within the $4$-dimensional spacetime. Furthermore, we can investigate the stability of this black hole in terms of the so-called Kaluza-Klein modes which are defined in the extra dimensions by computing QNMs, utilizing the perturbation equation we have derived~\cite{Cao:2021sty}. Given these findings, our goal is to investigate the impact of extra dimensions on QNMs within the framework of EGB gravity theory.

The high precision measurement also requires accurately calculating the QNMs. So far, the high precision methods have been made to calculate the frequencies of QNMs, such as the Wentzel-Kramers-Brillouin (WKB) approximation~\cite{Iyer:1986np,Konoplya:2003ii,Matyjasek:2017psv,Konoplya:2019hlu}, numerical integration method~\cite{Wang:2000dt,Wang:2004bv,Zhu:2001vi,Chabab:2017knz}, continued fractions method (CFM)~\cite{Leaver:1985ax,Leaver:1990zz,Nollert:1993zz,Konoplya:2004wg,Zhidenko:2006rs,Daghigh:2022uws}, asymptotic iteration method (AIM)~\cite{Cho:2009cj,Cho:2011sf,Mamani:2022akq} and so on. It is worth noting that the divergent behavior of the effective potential of asymptotic AdS spacetime we consider here will be quite different from that of asymptotic flat spacetime, and thus  lead to some subtleties for the QNM treatment. In this work, we use AIM  and numerical integration method jointly to do the calculation, two widely adopted methods which works well in the asymptotically AdS case with nonrational metric function. One can see some nice reviews~\cite{Konoplya:2011qq,Berti:2009kk,Zhao:2022lrl} to get more information about QNMs.  

The aim of this paper is to discuss characteristic mode frequencies of the tensor perturbation equation for the Maeda-Dadhich black hole obtained through the Kodama-Ishibashi formalism for general warped product spacetimes, with asymptotic iteration method and numerical integration method. After expressing the effect of extra dimensions with the help of the characteristic tensors of $\mathcal{K}^{n-4}$, we recast the perturbation equation derived in~\cite{Cao:2021sty} into a scalar field equation in four dimensional spacetime, which is quite different from the Klein-Gordon equation. Roughly speaking, the correction effect stems from the Gauss-Bonnet coupling constant $\alpha$ and the eigenvalues of characteristic tensors of the extra dimension part. The coefficients of the second order covariant derivative is no longer the spacetime metric. Actually, it is modified by the Einstein tensor of the four dimensional spacetime. In addition, there exists a term proportional to the scalar field in this equation. However, unlike the massive scalar field~\cite{Konoplya:2004wg,Zhidenko:2006rs}, the term is in fact dependent on the radial coordinate $r$. Having the scalar equation, we get the Schr\"{o}dinger-like equation by separating angular part as usual. Then, we calculate the QNMs by two different numerical methods, namely, asymptotic iteration method and numerical integration method. We then provide the characteristic frequencies  under different parameter choices, and study how these parameters affect the QNMs. It is found that the dimension of compactification part has no significant impact on the lifetime of QNMs.

The paper is organized as follows. In Sec. \ref{sec: Kaluza-Klein black hole}, we present a brief review of the Kaluza-Klein black hole proposed in~\cite{Maeda:2006iw}. The master equation for the perturbation of tensor-type in the Einstein–Gauss-Bonnet gravity theories is displayed in Sec. \ref{sec: master equation of tensor perturbation}. The Schr\"{o}dinger-like equation with the corresponding effective potential is also shown in the same section. In Sec. \ref{sec: AIM}, asymptotic iteration method is given to get the QNMs. In Sec. \ref{sec: time_domain_analysis}, the evolution of a scalar field is analysed. With the numerical results from Sec. \ref{sec: time_domain_analysis}, we use the Kumaresan-Tufts (KT) method to extract characteristic frequencies in Sec. \ref{sec: KT}. In Sec. \ref{sec: numerical_result_and_analysis}, a large amount of data will be displayed and we can find how the characteristic frequencies change with parameters. Section \ref{sec: conclusions_and_discussion} is devoted to conclusions and discussion. 

\section{Kaluza-Klein black hole}
\label{sec: Kaluza-Klein black hole}

In this section, we will have a brief review on the Kaluza-Klein black hole proposed by Maeda and Dadhich~\cite{Maeda:2006iw}. The action for $n\ge5$ in the $n$-dimensional spacetime with a metric $g_{MN}$ is given by 
\begin{eqnarray}
	S=\int \mathrm{d}^nx\sqrt{-g}\left[\frac{1}{2\kappa_n^2}\left(R-2\Lambda+\alpha L_{\text{GB}}\right)\right]+S_{\text{matter}}\, ,
\end{eqnarray}
where $\kappa_n$ is the coupling constant of gravity which depends on the dimension of spacetime, and $R$ and $\Lambda$ are the $n$-dimensional Ricci scalar and the cosmological constant, respectively. $S_{\text{matter}}$ stands for the matter fields. The Gauss-Bonnet term is given by 
\begin{equation}
	L_{\text{GB}}=R^2-4R_{MN}R^{MN}+R_{MNPQ}R^{MNPQ}\, ,
\end{equation}
where the capital letters $\{M,N,P,Q,\cdots\}$ are the indices for the $n$ dimensional spacetime. The symbol $\alpha$ is the coupling constant of the Gauss-Bonnet term. The symbol $\alpha$ is identified with the inverse string tension and is positive definite. The equation of motion of this theory is given by
\begin{eqnarray}\label{EOM}
	G_{MN}+\alpha H_{MN}+\Lambda g_{MN}=\kappa_n^2T_{MN}\, ,
\end{eqnarray}
where
\begin{eqnarray}
	G_{MN}=R_{MN}-\frac{1}{2}g_{MN}R\, ,
\end{eqnarray}
and
\begin{equation}
	\label{GB_tensor}
	H_{MN}=2\left[RR_{MN}-2R_{ML}R^{L}_{\ N}-2R^{KL}R_{MKNL}+R_{M}^{\ \ KLP}R_{NKLP}\right]-\frac{1}{2}g_{MN}L_{\text{GB}}\, .
\end{equation}

We consider the $n$-dimensional spacetime locally homeomorphic to $M^4\times\mathcal{K}^{n-4}$ with the metric, $g_{MN}=\text{diag}(g_{ab},r_0^2\gamma_{ij})$, where $a,b=0,\cdots,3$; $i,j=4,\cdots,n-1$. Here $g_{ab}$ is an arbitrary Lorentz metric on $M^4$, $r_0$ is a constant given by 
\begin{eqnarray}\label{r0}
	r_0^2=-2K\alpha(n-4)(n-5)\, ,
\end{eqnarray}
and $\gamma_{ij}$ is the unit metric on the $(n-4)$-dimensional space of constant curvature $\mathcal{K}^{n-4}$ with a sectional curvature $K=-1$.

We will seek a vacuum static solution with the metric on $M^4$ reading as
\begin{eqnarray}
	g_{ab}\mathrm{d}x^a\mathrm{d}x^b=-f(r)\mathrm{d}t^2+\frac{1}{f(r)}\mathrm{d}r^2+r^2\mathrm{d}\Sigma_{2(k)}^2\, ,
\end{eqnarray}
where $\mathrm{d}\Sigma_{2(k)}^2$ is the unit metric on two dimensional constant curvature space $\Sigma_{2(k)}$ with $k=\pm1, 0$. The governing equation is a single scalar equation on $M^4$, which is given by
\begin{eqnarray}\label{KK_main_equation}
	\frac{1}{n-4}{}^4\!{R}+\frac{\alpha}{2}{}^4\! L_{\text{GB}}+\frac{2n-11}{\alpha (n-4)^2(n-5)}=0\, ,
\end{eqnarray}
where ${}^4\!{R}$ and ${}^4\!L_{\text{GB}}$ are defined in the Lorentz manifold $(M^4, g_{ab})$. After some calculation, Eq. (\ref{KK_main_equation}) yields the general solution for the function $f(r)$: 
\begin{eqnarray}\label{f(r)}
	f(r)=k+\frac{r^2}{2(n-4)\alpha}\Biggl\{1\mp\Big[1-\frac{2n-11}{3(n-5)}+\frac{4(n-4)^2\alpha^{3/2}\mu}{r^3}-\frac{4(n-4)^2\alpha^2q}{r^4}\Big]^{1/2}\Biggr\}\, ,
\end{eqnarray}
where $\mu$ and $q$ are arbitrary dimensionless constants, and $\mu$ refers to the mass of the central object and $q$ is the charge-like parameter. Probably, due to the topology of the spacetime, i.e., $\mathcal{M}^n \simeq M^4 \times \mathcal{K}^{n-4}$ with constant curvature $K=-1$, the charge parameter $q$ automatically appears as a constant of integration. It should be noted this kind of charge corresponds to the so-called ``Weyl charge" defined by the integration of the Weyl tensor projected onto the brane~\cite{Shiromizu:1999wj}. Detailed explanation on the meaning of this charge can be found in~\cite{Maeda:2006iw} and references therein.

To make $f(r)$ meaningful, the dimension of spacetime must be set as $n\ge6$. There are two branches of the solution indicated by a sign in front of the square root in Eq. (\ref{f(r)}), which we call the minus and the plus branch~\cite{Maeda:2006iw}. We will focus on the case with $k=1$ in the following sections. Since the expression in the radical of the metric function should be non-negative, the parameter $\mu$ and $q$ should meet the following condition
\begin{eqnarray}\label{condition_alpha_mu}
	1-\frac{2n-11}{3(n-5)}+\frac{4(n-4)^2\alpha^{3/2}\mu}{r^3}-\frac{4(n-4)^2\alpha^2q}{r^4}\ge0\, .
\end{eqnarray}
A sufficient condition is that $\mu\ge0$ and $q\le0$, and then $r\in(0,+\infty)$. Notice that since only for the negative branch, the metric function $f(r)$ may have zero points, so we choose the negative branch for our study for the black hole appearing, i.e., 
\begin{eqnarray}\label{f(r)_k_1}
	f(r)=1+\frac{r^2}{2(n-4)\alpha}\Biggl\{1-\Big[1-\frac{2n-11}{3(n-5)}+\frac{4(n-4)^2\alpha^{3/2}\mu}{r^3}-\frac{4(n-4)^2\alpha^2q}{r^4}\Big]^{1/2}\Biggr\}\, .
\end{eqnarray}
The function $f(r)$ is expanded for $r\to+\infty$ as
\begin{eqnarray}
	f(r)&=&1+\frac{r^2}{2(n-4)\alpha}\Big[1-\sqrt{\frac{n-4}{3(n-5)}}\Big]\nonumber\\
	&&-\frac{\alpha^{1/2}\mu\sqrt{3(n-4)(n-5)}}{r}+\frac{\alpha q\sqrt{3(n-4)(n-5)}}{r^2}+\mathcal{O}\Big(\frac{1}{r^3}\Big)\, .
\end{eqnarray}
This is the same as the Reissner-Nordstrom-anti-de Sitter (RNAdS) spacetime for $k=1$ in spite of the absence of the Maxwell field. 

Since the design of the algorithm of QNMs involves the number of the zero points of the metric function $f(r)$, we will find the number of the zero points of $f(r)$. $f(r)=0$ is equivalent to $h(r)=0$ in terms of the condition $\mu\ge0$ and $q\le0$ where 
\begin{eqnarray}
	h(r)=\frac{2n-11}{12(n-5)(n-4)^2\alpha^2}r^4+\frac{r^2}{(n-4)\alpha}-\frac{\mu}{\alpha^{1/2}}r+(q+1)\, .
\end{eqnarray}
The derivative of $h(r)$ is 
\begin{eqnarray}
	h^{\prime}(r)=\frac{2n-11}{3(n-5)(n-4)^2\alpha^2}r^3+\frac{2}{(n-4)\alpha}r-\frac{\mu}{\alpha^{1/2}}\, .
\end{eqnarray}
It is easy to find that $h^{\prime}(r)$ only has one zero point in $(0,+\infty)$. Therefore, $f(r)$ has two zero points at most. The ranges of parameter values of $\mu$ and $q$ are selected as $\mu\ge0$ and $q\le0$ for simplicity. Additionally, we can easily see that when $q<-1$, $f(r)$ has and only has one zero point, i.e., only the event horizon exists. In later calculation, we can judge whether it has one or two zero points through numerical calculations. The event horizon is denoted as $r_{+}$ and the inner horizon is denoted as $r_{-}$ if it exists.

\section{the master equation of the tensor perturbation}
\label{sec: master equation of tensor perturbation}

We consider a $n=4+(n-4)$ dimensional spacetime $(\mathcal{M}^n,g_{MN})$, which has a local direct product manifold with a metric
\begin{eqnarray}
g_{MN}\mathrm{d}x^M\mathrm{d}x^N=g_{ab}(y)\mathrm{d}y^a\mathrm{d}y^b+r^2(y)\gamma_{ij}(z)\mathrm{d}z^i\mathrm{d}z^j\, ,
\end{eqnarray}
where coordinates $x^M=\{y^1,\cdots,y^4;z^1,\cdots,z^{(n-4)}\}$. In the following discussion, the Riemann manifold $(\mathcal{N}^{n-4},\gamma_{ij})$ is assumed to be a maximally symmetric space, i.e., $\mathcal{N}^{n-4}=\mathcal{K}^{n-4}$. The metric compatible covariant derivatives associated with $g_{ab}$ and $\gamma_{ij}$ are denoted by $D_a$ and $\hat{D}_i$. $K$ is the sectional curvature of the space and for this Kaluza-Klein black hole we have $K=-1$. 

Under the linear perturbation of the metric $g_{MN}\to g_{MN}+h_{MN}$, the linear perturbation equations for Eq.(\ref{EOM}) can be obtained. The tensor perturbation equation is obtained by making the metric perturbation $h_{MN}$ and the energy-momentum perturbation $\delta T_{MN}$ meet with
\begin{eqnarray}
    h_{ab}=0\, ,\quad h_{ai}=0\, ,\nonumber\\
    \delta T_{ab}=0\, ,\quad \delta T_{ai}=0\, .
\end{eqnarray}
At the same time, the Riemann part of the perturbation $h_{MN}$, i.e., $h_{ij}$, is transverse and traceless. The scalar-type and vector-type perturbation equations are the equations for other parts of $h_{MN}$~\cite{Ishibashi:2011ws,Cai:2013cja}. After some calculation, the master equation of tensor perturbation (The computing method can be found in~\cite{Cai:2013cja,Cao:2021sty}.) in vacuum can be written as~\cite{Cao:2021sty}
\begin{eqnarray}\label{master_equation} (P^{ab}D_aD_b+P^{mn}\hat{D}_m\hat{D}_n+P^aD_a+V)\Big(\frac{h_{ij}}{r^2}\Big)=0\, ,
\end{eqnarray}
where
\begin{eqnarray}\label{Pab}
	P^{ab}= g^{ab} + 2(n-6) \alpha\left\lbrace 2\frac{D^aD^br}{r}+\left[(n-7)\frac{K-(Dr)^2}{r^2}-2\frac{\prescript{4}{}{\Box}r}{r}\right]g^{ab}\right\rbrace-4\alpha\cdot{}^4\! G^{ab}\, ,
\end{eqnarray}
\begin{eqnarray}\label{Pmn}
	P^{mn}=\Bigg\{1+2 \alpha\left[ {}^4\!{R}- \frac{2(n-7) {}^4\!{\Box}r}{r}+ (n-7)(n-8)\frac{K-(Dr)^2}{r^2}\right] \Bigg\}\frac{\gamma^{mn }}{r^2}\equiv\frac{Q}{r^2}\gamma^{mn}\, ,
\end{eqnarray}
\begin{eqnarray}\label{Pa}
	P^{a}&=&(n-4)\frac{D^ar}{r} +  2(n-6) \alpha \Bigg\{4\frac{D^aD^br}{r}+\Big[ {}^4\!{R}
	-2(n-5)\frac{{}^4\!{\Box}r}{r}\nonumber\\ 
	&&+(n-6)(n-7)\frac{K-(Dr)^2}{r^2}\Big]g^{ab}\Bigg\}\frac{D_br}{r}- 8\alpha\cdot {}^4\! G^{ab}\frac{D_br}{r}\, ,
\end{eqnarray}
and
\begin{eqnarray}\label{V}
	V&=&{}^4\!{R}-2(n-5)\frac{{}^4\! {\Box}r}{r}+\frac{(n-4)(n-7)K}{r^2}-\frac{(n-5)(n-6)(Dr)^2}{r^2}-2\Lambda\nonumber\\
	&&+\alpha\Bigg\{{}^4\! L_{\text{GB}}+8(n-5)\cdot {}^4\! G^{ab}\frac{D_aD_br}{r} -4(n-5)(n-6)\frac{(D^aD^br)(D_aD_br)}{r^2}\nonumber\\
	&&+4(n-5)(n-6)\left(\frac{{}^4\! {\Box}r}{r}\right)^2+2(n-4)(n-7)\frac{K\cdot {}^4\! R}{r^2}-2(n-5)(n-6)\frac{(Dr)^2\cdot {}^4\! {R}}{r^2}\nonumber\\
	&&-4(n-4)(n-7)^2\frac{K\cdot {}^4\! {\Box}r}{r^3}+4(n-5)(n-6)(n-7)\frac{(Dr)^2\cdot{}^4\! {\Box}r}{r^3}\nonumber\\
	&&-2(n-4)(n-7)^2(n-8)\frac{K\cdot(Dr)^2}{r^4}+(n-7)(n-8)[(n-4)^2-3(n-4)-2]\frac{K^2}{r^4}\nonumber\\
	&&+(n-5)(n-6)(n-7)(n-8)\left[\frac{(Dr)^2}{r^2}\right]^2\Bigg{\}}\, ,
\end{eqnarray}
where ${}^4\!\square=g^{ab}D_aD_b$ is the d'Alembertian in $(M^4,g_{ab})$. 
We can apply the separation of variables~\cite{Dotti:2004sh}, 
\begin{eqnarray}\label{separate_variables}
	h_{ij}(y,z^1,\cdots,z^{n-4})=r^2\Phi(y)\bar{h}_{ij}(z^1,\cdots,z^{n-4})\, ,
\end{eqnarray}
where $\bar{h}_{ij}$ is the characteristic tensor of $\mathcal{K}^{n-4}$ and satisfies,
\begin{eqnarray}\label{gamma}
	\hat{D}^k\hat{D}_k\bar{h}_{ij}=\gamma\bar{h}_{ij}\, ,\quad\hat{D}^i\bar{h}_{ij}=0\, ,\quad\gamma^{ij}\bar{h}_{ij}=0\, .
\end{eqnarray}
Then, one obtains a four dimensional wave equation of $\Phi$ on the manifold $M^4$ as follows (One can find the details in Appendix.\ref{app_1}.)
\begin{eqnarray}\label{master_equation_4d}
    \Big[\frac{4n-22}{(n-4)(n-5)}g^{ab}-4\alpha\cdot{}^4\! G^{ab}\Big]D_aD_b\Phi+\Big[\frac{2+\gamma}{(n-4)(n-5)}{}^4\!{R}+\frac{3(n-6)(2+\gamma)}{\alpha(n-4)^2(n-5)^2}\Big]\Phi=0\, .
\end{eqnarray}
The equation \eqref{master_equation} we initially obtained is a tensor equation about $h_{ij}$, which turns into this scalar equation on the manifold $M^4$ after separating the variables. In fact, the scalar $\Phi$ is the amplitude of the characteristic field $\bar{h}_{ij}$ corresponding to the characteristic value $\gamma$. Comparing with the standard Klein-Gordon equation ${}^4\!\square\Phi=0$ in $(M^4\, ,g_{ab})$, it can be found that the coefficient of the second derivative  of $\Phi$ in Eq. (\ref{master_equation_4d}) is added by a term related to four dimensional Einstein tensor $G^{ab}$.

Solutions to equations (\ref{gamma}) are worked out in~\cite{Higuchi:1986wu} for $K=1$, where it is shown that the spectrum of eigenvalues is $\gamma=-L(L+n-5)+2\, ,L=2,3,4,\cdots$. However, as for our case $K=-1$, there is a subtlety in the value of $\gamma$ which may be a difficult mathematical problem. To avoid mathematical hardship, as a matter of convenience, $\gamma\in\mathbb{R}$ is assumed. It can be seen that QNMs can be obtained for a given $\gamma$.
Upon examination of Eqs. (\ref{gamma}), it can be observed that a total of $n-3$ constraints are imposed on $\bar{h}_{ij}$. The degrees of freedom for $\bar{h}_{ij}$ are $(n-4)(n-3)/2$. In the specific case where $n=6$, the number of constraints imposed on $\bar{h}_{ij}$ is equal to its degrees of freedom. As a result, $\bar{h}_{ij}$ possesses no degrees of freedom for propagation. Since the issue mentioned above is excluded when $n\ge7$, we will focus on the cases where $n\ge7$ for the remainder of this paper.

Now, Eq. (\ref{master_equation_4d}) is an equation about $\Phi$ on $M^4$ with a Lorentz metric 
\begin{eqnarray}
	g_{ab}\mathrm{d}x^a\mathrm{d}x^b=-f(r)\mathrm{d}t^2+\frac{1}{f(r)}\mathrm{d}r^2+r^2(\mathrm{d}\theta^2+\sin^2\theta\mathrm{d}\phi^2)\, ,
\end{eqnarray} 
where we have chosen the metric function (\ref{f(r)}) with $k=1$. Separating the variables as
\begin{eqnarray}
	\Phi(t,r,\theta,\phi)=e^{-i\omega t}R(r)Y(\theta,\phi)\, ,
\end{eqnarray}
where $Y(\theta,\phi)$ is the spherical harmonics, we get the radial equation of $R(r)$ as follows,
\begin{eqnarray}
	R^{\prime\prime}+B(r)R^{\prime}+C(r)R=0\, ,\label{radial_equation}
\end{eqnarray}
where the functions $B(r)$ and $C(r)$ are
\begin{eqnarray}\label{B(r)}
	B(r)&=&\Big[\frac{4n-22}{(n-4)(n-5)}\Big(f^{\prime}+\frac{2f}{r}\Big)-4\alpha\frac{-f^{\prime}+3ff^{\prime}+r(f^{\prime})^2+rff^{\prime\prime}}{r^2}\Big]\nonumber\\
    &&\times\Big[\frac{4n-22}{(n-4)(n-5)}f-\frac{4\alpha f(-1+f+rf^{\prime})}{r^2}\Big]^{-1}\, ,
\end{eqnarray}
and
\begin{eqnarray}\label{C(r)}
	C(r)&=&\frac{\omega^2}{f^2}+\Biggl\{-\frac{l(l+1)}{r^2}\Big[\frac{4n-22}{(n-4)(n-5)}-\frac{2\alpha(2f^{\prime}+rf^{\prime\prime})}{r}\Big]\nonumber\\
	&&+\Big[\frac{2+\gamma}{(n-4)(n-5)}{}^4\!{R}+\frac{3(n-6)(2+\gamma)}{\alpha(n-4)^2(n-5)^2}\Big]\Biggr\}\Big[\frac{4n-22}{(n-4)(n-5)}f-\frac{4\alpha f(-1+f+rf^{\prime})}{r^2}\Big]^{-1}\, .
\end{eqnarray}
Here, the $``~\prime~"$ denotes the derivative with respect to $r$. 

Now, our task is to bring this equation to the more familiar form of one-dimensional Schr\"{o}dinger-like equation, for which we need to remove the friction term in the equation above. There are two transformations that we can consider: a change of variable for the radial coordinate $r_{\star}=r_{\star}(r)$ and a rescaling of $R$, so that 
\begin{eqnarray}\label{transformations}
	\mathrm{d}r_{\star}=z(r)\mathrm{d}r\, ,\quad R=S(r)\varphi\, , 
\end{eqnarray} 
for given functions $S$ and $z$, and $\varphi$ is the new radial function now~\cite{Cano:2020cao}. Performing the transformations (\ref{transformations}), one gets 
\begin{eqnarray}\label{master_equation_tortoise}
	z^2S\frac{\mathrm{d}^2\varphi}{\mathrm{d}r_{\star}^2}+\Big(z^{\prime}S+2zS^{\prime}+BzS\Big)\frac{\mathrm{d}\varphi}{\mathrm{d}r_{\star}}+\Big(CS+BS^{\prime}+S^{\prime\prime}\Big)\varphi=0\, .
\end{eqnarray}
In order to remove the term $\mathrm{d}\varphi/\mathrm{d}r_{\star}$, it is found that $S$ and $z$ must satisfy
\begin{eqnarray}\label{relation}
	zS^2=\exp\Big(-\int B(r)\mathrm{d}r\Big)\, .
\end{eqnarray}
Therefore, Eq. (\ref{master_equation_tortoise}) becomes
\begin{eqnarray}
	\frac{\mathrm{d}^2\varphi}{\mathrm{d}r_{\star}^2}+\frac{CS+BS^{\prime}+S^{\prime\prime}}{z^2S}\varphi=0\, .
\end{eqnarray}
We have the freedom to fix one of these functions, the other one will then be determined by the relation (\ref{relation}). We make the following choice of tortoise coordinate:
\begin{eqnarray}
	z(r)=\frac{1}{f(r)}\, .
\end{eqnarray}
It is found that $r_\star$ is finite as $r\to+\infty$, and $r_\star\to-\infty$ as $r\to r_{+}$. This is similar to the AdS case. The function $S$ is satisfied with
\begin{eqnarray}
	S^2&=&\frac{1}{z}\exp\Big(-\int B\mathrm{d}r\Big)\, ,\nonumber\\
	(S^2)^{\prime}&=&\frac{-Bz-z^{\prime}}{z^2}\exp\Big(-\int B\mathrm{d}r\Big)\, ,\nonumber\\
	(S^2)^{\prime\prime}&=&\Big[-\frac{z^{\prime\prime}}{z^2}+\frac{2Bz^{\prime}}{z^2}+\frac{2(z^{\prime})^2}{z^3}-\frac{B^{\prime}}{z}+\frac{B^2}{z}\Big]\exp\Big(-\int B\mathrm{d}r\Big)\, .
\end{eqnarray}
Finally, the standard Schr\"{o}dinger-like equation is obtained 
\begin{eqnarray}\label{Schrodinger_equation}
	\Big[\frac{\mathrm{d}^2}{\mathrm{d}r_{\star}^2}+(\omega^2-V_{\text{eff}})\Big]\varphi=0\, ,
\end{eqnarray}
where the effective potential $V_{\text{eff}}$ is
\begin{eqnarray}\label{effective_potential}
	V_{\text{eff}}&=&\omega^2-\frac{CS+BS^{\prime}+S^{\prime\prime}}{z^2S}\nonumber\\
	&=&\omega^2-f^2C+\frac{(f^{\prime})^2}{4}-\frac{ff^{\prime\prime}}{2}+\frac{f^2B^{\prime}}{2}+\frac{f^2B^2}{4}\, .
\end{eqnarray} 
It should be noted that $V_{\text{eff}}$ above is independent of $\omega$ since there is a term $\omega^2/f^2$ in $C$.

We need to examine the behavior of the effective potential $V_{\text{eff}}$ before calculating QNMs. There are two things we must accomplish. 
First, we should check whether there exists any $r=r_V\in(r_{+},+\infty)$ such that $V_{\text{eff}}$ is divergent, i.e., 
\begin{eqnarray}\label{divergent_V}
	\lim_{r\to r_V}V_{\text{eff}}=\infty\, .
\end{eqnarray}
In other words, we are supposed to pay attention to whether there exists $r=r_V\in(r_{+},+\infty)$ such that 
\begin{eqnarray}
	\frac{4n-22}{(n-4)(n-5)}-\frac{4\alpha(-1+f+rf^{\prime})}{r^2}=0\, ,
\end{eqnarray}
which is is equivalent to
\begin{eqnarray}
	&&\frac{2n-11}{n-5}r^8+12\alpha^{3/2}\mu(n-4)(2n-11)r^5-12\alpha^2q(n-4)(n-6)r^4\nonumber\\
	&&-144\alpha^{7/2}\mu q(n-5)^2(n-4)^2r+48\alpha^4(n-5)^2(n-4)^2q^2=0\, .
\end{eqnarray}
However, in the above polynomial equation, we find each term is nonnegative provided that $\mu\ge0$, $q\le0$ and $n\ge6$, i.e., there is no such $r_V$ that Eq. (\ref{divergent_V}) is established. Hence, the effective potential $V_{\text{eff}}$ is always regular at $(r_{+},+\infty)$. Second, we should acquaint ourselves with the asymptotic behavior of $V_{\text{eff}}$ as $r\to+\infty$ and $r\to r_{+}$. When $r\to+\infty$, we have
\begin{eqnarray}
    \label{eq: effective_potential_spatial_infinity}
	V_{\text{eff}}&=&\Biggl\{\frac{(\gamma +2) \Big(2 \sqrt{3}\sqrt{\frac{n-4}{n-5}} n-3 n-10 \sqrt{3} \sqrt{\frac{n-4}{n-5}}+12\Big) \Big(\sqrt{3}\sqrt{\frac{n-4}{n-5}}-3\Big)}{12 \alpha ^2 (n-5) (n-4)^2 \Big(\sqrt{3} \sqrt{\frac{n-4}{n-5}} n-n-5 \sqrt{3} \sqrt{\frac{n-4}{n-5}}+4\Big)}+\frac{\Big(\sqrt{3} \sqrt{\frac{n-4}{n-5}}-3\Big)^2}{18 \alpha^2(n-4)^2}\Biggr\}r^2+\mathcal{O}(1)\nonumber\\
	&=&\Big(3-\sqrt{3}\sqrt{\frac{n-4}{n-5}}\Big)\Biggl\{36\alpha^2(n-5)(n-4)^2\Big[\sqrt{3}\sqrt{\frac{n-4}{n-5}}(n-5)-(n-4)\Big]\Biggr\}^{-1}\times\nonumber\\
	&&\Biggl\{\Big[9(n-4)-6\sqrt{3}(n-5)\sqrt{\frac{n-4}{n-5}}\Big]\gamma+2\sqrt{3}\sqrt{\frac{n-4}{n-5}}(n-5)(4n-25)-6(n-4)(2n-13)\Biggr\}r^2+\mathcal{O}(1)\nonumber\nonumber\\
	&\equiv&V_0(\alpha,n,\gamma)r^2+\mathcal{O}(1)\, .
\end{eqnarray}
The stability requirement demands that $V_{\text{eff}}$ tends towards positive infinity, i.e., $V_0(\alpha,n,\gamma)>0$~\cite{1995AmJPh63256B}. One can find the range of $\gamma$ in terms of the dimension of spacetime $n$ in Tab.\ref{tab1}. An important property of the effective potential  $V_{\text{eff}}$ is that 
\begin{eqnarray}\label{effective_potential_event_horizon}
	V_{\text{eff}}(r_{+})=0\, .
\end{eqnarray}
One can check it by a direct calculation (see Appendix.\ref{app_2}).

\begin{table}[h]
\begin{ruledtabular}
\caption{The range of $\gamma$ in terms of the dimension of spacetime $n$}
\label{tab1}
\begin{tabular}{cc}
the dimension of spacetime $n$ & the range of $\gamma$ \\
\colrule
  $n=7$& $\gamma>-\frac{2\sqrt{3}\sqrt{\frac{n-4}{n-5}}(n-5)(4n-25)-6(n-4)(2n-13)}{9(n-4)-6\sqrt{3}(n-5)\sqrt{\frac{n-4}{n-5}}}$ \\
         $n=8$& $\mathbb{R}$ \\
         $n\ge9$& $\gamma<-\frac{2\sqrt{3}\sqrt{\frac{n-4}{n-5}}(n-5)(4n-25)-6(n-4)(2n-13)}{9(n-4)-6\sqrt{3}(n-5)\sqrt{\frac{n-4}{n-5}}}$\\

\end{tabular}
\end{ruledtabular}
\end{table}

Before proceeding to actual numerical calculation, we make some remarks on our analysis. Although Newton's gravitational constant varies with different total dimensions of spacetime, it does not directly enter into the calculation, since the energy-momentum tensor vanishes in terms of the considered spacetime. We utilized Newton's gravitational constant only when converting mass dimensions to length dimensions. As a result, we can simply set both the speed of light $c$ and the $4$-dimensional Newton's gravitational constant $G$ to $1$ without encountering any issues with the higher-dimensional Newton's gravitational constant. The Gauss-Bonnet coupling, denoted by $\alpha$, has the dimension of length squared. Due to the asymptotic property of the Kaluza-Klein black hole we studied, the effective AdS length is~\cite{Maeda:2006iw} 
\begin{eqnarray}
l_{\text{eff}}=\Bigg[\dfrac{2(n-4)\alpha}{1-\sqrt{\dfrac{n-4}{3(n-5)}}}\Bigg]^{1/2} \, ,
\end{eqnarray}
the effective mass of the black hole is represented by
\begin{eqnarray}
    M_{\text{eff}}=\dfrac{1}{2}\alpha^{1/2}\mu\sqrt{3(n-4)(n-5)}\, ,
\end{eqnarray}
while the effective charge of the black hole is 
\begin{eqnarray}
    Q_{\text{eff}}=\Big(\alpha q\sqrt{3(n-4)(n-5)}\Big)^{1/2}\, .
\end{eqnarray}
The mass parameter $\mu$, the charge-like parameter $q$ are both dimensionless and the Gauss-Bonnet coupling constant $\alpha$ is the only independent length scale, i.e., other length scales can be determined by $\alpha$. The quasi-normal frequencies, denoted by $\omega$, have the dimension of length inversed, which means that the quantity $\sqrt{\alpha} \omega$ is dimensionless. Therefore, as the value of $\alpha$ is changed, it merely results in a scaling transformation, leaving the dimensionless quantities unaffected, i.e., the quantity $\sqrt{\alpha} \omega$ remains constant when only the value of $\alpha$ varies. One can prove it by using a scale transformation for the Schr\"{o}dinger-like equation (\ref{Schrodinger_equation}) and this conclusion is also verified numerically (see Tab. \ref{tab: AIM_alpha}). We omit the unit of $\alpha$, but one can introduce the unit of $\alpha$ as $\mathrm{s}^2$. Consequently, the unit of $\omega$ will be adjusted to hertz.

\checked{}

\section{Asymptotic iteration method}\label{sec: AIM}
In this section, we will use the asymptotic iteration method (AIM) to solve the QNMs of the tensor perturbation in Kaluza-Klein black hole for EGB gravity. At the beginning, we provide a brief review on the asymptotic iteration method. Consider a second order homogeneous linear differential equation for the function $\chi(x)$,
\begin{eqnarray}\label{second_ode}
    \chi^{\prime\prime}(x)=\lambda_0(x)\chi^{\prime}(x)+s_0(x)\chi(x)\, ,
\end{eqnarray}
where $\lambda_0(x)\neq0$. Differentiating Eq. (\ref{second_ode}) with respect to $x$, one finds
\begin{eqnarray}
    \chi^{\prime\prime\prime}(x)=\lambda_1(x)\chi^{\prime}(x)+s_1(x)\chi(x)\, ,
\end{eqnarray}
where
\begin{eqnarray}
    \lambda_1(x)&=&\lambda_0^{\prime}+s_0+\lambda_0^2\, ,\nonumber\\
    s_1(x)&=&s^{\prime}_0+s_0\lambda_0\, .
\end{eqnarray}
Iteratively, the $(\bar{n}-1)$-th and $\bar{n}$-th differentiations of Eq. (\ref{second_ode}) give
\begin{eqnarray}
    \chi^{(\bar{n}+1)}&=&\lambda_{\bar{n}-1}(x)\chi^{\prime}(x)+s_{\bar{n}-1}\chi(x)\, ,\nonumber\\
    \chi^{(\bar{n}+2)}&=&\lambda_{\bar{n}}(x)\chi^{\prime}(x)+s_{\bar{n}}(x)\chi(x)\, ,
\end{eqnarray}
where
\begin{eqnarray}
    \lambda_{\bar{n}}(x)&=&\lambda^{\prime}_{\bar{n}-1}+s_{\bar{n}-1}+\lambda_0\lambda_{\bar{n}-1}\, ,\nonumber\\
    s_{\bar{n}}(x)&=&s^{\prime}_{\bar{n}-1}+s_0\lambda_{\bar{n}-1}\, .
\end{eqnarray}
The so-called  ``quantization condition" is given by
\begin{eqnarray}\label{quantization_condition}
    s_{\bar{n}}(x)\lambda_{\bar{n}-1}(x)-s_{\bar{n}-1}(x)\lambda_{\bar{n}}(x)=0\, .
\end{eqnarray}
It is noted that at each iteration one must take the derivative of the $s$ and $\lambda$ terms of the previous iteration~\cite{Ciftci:2005xn}. This ``deficiency" might bring difficulties for numerical calculations. An improved version of the AIM which bypasses the need to take derivatives at each step is proposed in~\cite{Cho:2009cj,Cho:2011sf}. This greatly improves both the accuracy and speed of the method. The functions $\lambda_{\bar{n}}$ and $s_{\bar{n}}$ are expanded in a Taylor series around the point $\xi_0$ at which the AIM is performed, which means that
\begin{eqnarray}
	\lambda_{\bar{n}}(\xi)=\sum_{j=0}^{\infty}c^j_{\bar{n}}(\xi-\xi_0)^j\, ,
    \\
	s_{\bar{n}}(\xi)=\sum_{j=0}^{\infty}d^j_{\bar{n}}(\xi-\xi_0)^j\, ,
\end{eqnarray}
where $c^j_{\bar{n}}$ and $d^j_{\bar{n}}$ are the $j$-th Taylor coefficients of $\lambda_{\bar{n}}(\xi)$ and $s_{\bar{n}}(\xi)$ respectively. Substituting these expressions, we get a set of recursion relations for the coefficients:
\begin{eqnarray}\label{recursion_relations}
	c^j_{\bar{n}}&=&(j+1)c^{j+1}_{\bar{n}-1}+d^j_{\bar{n}-1}+\sum_{k=0}^{j}c_0^kc_{\bar{n}-1}^{j-k}\, ,\nonumber\\
	d^i_{\bar{n}}&=&(j+1)d^{j+1}_{\bar{n}-1}+\sum_{k=0}^{j}d^k_0c^{j-k}_{\bar{n}-1}\, .
\end{eqnarray}
In terms of these coefficients, the ``quantization condition" (\ref{quantization_condition}) can be expressed as
\begin{eqnarray}\label{quantization_condition_AIM}
	d^0_{\bar{n}}c^0_{\bar{n}-1}-d^0_{\bar{n}-1}c^0_{\bar{n}}=0\, .
\end{eqnarray}
Thus we have reduced the AIM into a set of recursion relations which no longer require derivative operations. The symbol $\bar{n}$ is nothing but the iteration order.

Now, we can find the standard AIM form of Schr\"{o}dinger-like equation (\ref{Schrodinger_equation}) with the help of the boundary conditions of QNMs. The effective potential is zero at the event horizon $r\to r_{+}$ [see Eq. (\ref{effective_potential_event_horizon})]. The Dirichlet boundary condition is added at spatial infinity because of the divergence of the effective potential there, as we can see through Eq. (\ref{eq: effective_potential_spatial_infinity}). Therefore, the boundary conditions are taken so that the asymptotic behavior of the solutions is
\begin{eqnarray}\label{boundary_condition}
	\varphi(r)\to\left\{
	\begin{aligned}
	&e^{-i\omega r_{\star}}\quad &r\to r_{+}\, ,\\
        &0\quad &r\to+\infty\, ,
	\end{aligned}
	\right. 
    \end{eqnarray}
which represents an in-going wave at the event horizon and no waves at infinity. According to the theory of second order ordinary differential equation, we know that $r=r_{+}$ and $r=+\infty$ both are regular singular points.  In order to apply the boundary condition (\ref{boundary_condition}), we define the following solution (One can see more details in Appendix \ref{app_3}.)
\begin{eqnarray}
	\varphi(r)=\Big(\frac{r-r_{+}}{r-r_{-}}\Big)^{-i\omega/f^{\prime}(r_{+})}\Big(\frac{r_{+}-r_{-}}{r-r_{-}}\Big)^{\rho}\tilde{\varphi}(r)\, ,
\end{eqnarray}
in which the index at infinity is (we have disposed another index due to the boundary condition of $\varphi(r)$ at infinity.)
\begin{eqnarray}
	\rho=\frac{1}{2}\Bigg{\{}1+\sqrt{1+\frac{16(n-4)^2\alpha^2V_0}{\Big[1-\sqrt{\frac{n-4}{3(n-5)}}\Big]^2}}\Bigg{\}}>0\, ,
\end{eqnarray}
and $\tilde{\varphi}(r)$ is a finite and convergent function. Based on the above discussion, a compact coordinate is introduced as follows
\begin{eqnarray}\label{compact_coordinate}
	\xi=\frac{r-r_{+}}{r-r_{-}}\, ,
\end{eqnarray}
with $0\le\xi<1$. If there exists only the event horizon $r_{+}$, we will set $r_{-}=0$ in Eq. (\ref{compact_coordinate}). The regular function $\tilde{\varphi}(\xi)$ is introduced as 
\begin{eqnarray}\label{regular_function}
	\varphi(\xi)=\xi^{-i\omega/f^{\prime}(r_{+})}(1-\xi)^\rho\tilde{\varphi}(\xi)\, ,
\end{eqnarray}
so that the function $\varphi(\xi)$ obeys the Dirichlet boundary condition at spatial infinity $(\xi=1)$. Now, we will rewrite the Eq. (\ref{Schrodinger_equation}) into the differential equation for the regular function $\tilde{\varphi}(\xi)$ by using Eq. (\ref{compact_coordinate}) and Eq. (\ref{regular_function}). Using the relationship (\ref{compact_coordinate}) between $\xi$ and $r$, we can derive the inverse relationship between $\xi$ and $r$, which is expressed as
\begin{eqnarray}
	r=\frac{r_{+}-\xi r_{-}}{1-\xi}\, .
\end{eqnarray}
After some calculations, the standard AIM form of Eq.(\ref{Schrodinger_equation}) equipped with the boundary condition (\ref{boundary_condition}) is found as below 
\begin{eqnarray}
	\frac{\mathrm{d}^2\tilde{\varphi}}{\mathrm{d}\xi^2}=\lambda_0(\xi)\frac{\mathrm{d}\tilde{\varphi}}{\mathrm{d}\xi}+s_0(\xi)\tilde{\varphi}\, ,
\end{eqnarray}
where
\begin{eqnarray}
	\lambda_0(\xi)=-\frac{2\kappa\xi(\rho +1)-i(\xi-1)\omega}{\kappa(\xi-1)\xi}-\frac{g(\xi ) (r_{+}-r_{-})}{(\xi -1)^2 f(\xi )}\, ,
\end{eqnarray}
and
\begin{eqnarray}
	s_0(\xi)&=&\frac{1}{4\kappa^2(\xi-1)^4\xi^2f(\xi)^2}\Biggl\{-2 \kappa(\xi-1)\xi  f(\xi) g(\xi) (r_{+}-r_{-}) \Big[2\kappa\xi\rho-i(\xi-1)\omega\Big]\nonumber\\
	&&+(\xi-1)^2 f^2(\xi)\Big[-4\kappa^2\xi^2\rho(\rho+1)+2i\kappa(\xi-1)\omega(2 \xi\rho+\xi+1)+(\xi-1)^2\omega^2\Big]\nonumber\\
	&&-4\kappa^2\xi^2 (r_{+}-r_{-})^2\Big[\omega ^2-V_{\text{eff}}(\xi)\Big]\Biggr\}\, ,
\end{eqnarray}
with $\kappa=f^{\prime}(r_{+})/2$ and
\begin{eqnarray}
	g(\xi)\equiv f^{\prime}(r)|_{r=(r_{+}-\xi r_{-})/(1-\xi)}\, .
\end{eqnarray}
These equations are now in the standard form for AIM calculation, and we can use the standard AIM treatment to derive the QNM frequencies.

The QNM frequencies depend on six physical parameters, namely, the spacetime dimension $n$, the coupling constant $\alpha$, the black hole mass parameter $\mu$, the black hole charge parameter $q$, and the ``quantum numbers" $l$ and $\gamma$. We will demonstrate how these six parameters influence the QNM frequencies later. Besides, the numerical results also depend on two nonphysical parameters, the iteration order and the expanding position $\xi_0$ in AIM. Before going into the discussion about physical parameter, we will first give a discussion about these two parameters.

\begin{figure}[htbp]
	\centering
	\includegraphics[width=0.6\textwidth]{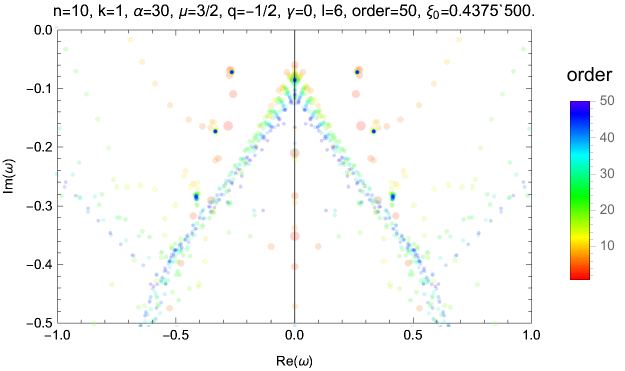}
	\caption{The AIM numerical results of iteration order from $1$st to $50$th. The parameter choice is as shown in the figure. The horizontal axis stands for the real part of frequencies, while the vertical axis stands for the imaginary part of frequencies. Points of different color are from different iteration orders, as indicated by the legend bar on the right. Since we expect the physical frequencies repeatedly appear in the solution of each iteration order, we may recognize the common position of various different colors to be the location of physical frequencies, while those points which show only one color are considered as numerical artifacts.}
	\label{fig: AIM_order}
\end{figure}

In Fig. \ref{fig: AIM_order}, we illustrate how the iteration order influences the numerical result. This figure shows the numerical results for iteration orders ranging from 1st to 50th, with the parameter choices indicated in the figure. The colors of the points correspond to the iteration order, as shown in the legend bar. As the iteration order increases, the physical frequencies, which are the physical roots of Eq. (\ref{quantization_condition_AIM}), repeatedly appear in the results of each iteration order. Therefore, we may recognize physical frequencies through the common locations of points of different colors, while those points with only one color are from numerical artifacts. We expect that the precision of the numerical result will also increase with increasing iteration order.

\begin{figure}[htbp]
	\centering
	\includegraphics[width=0.45\textwidth]{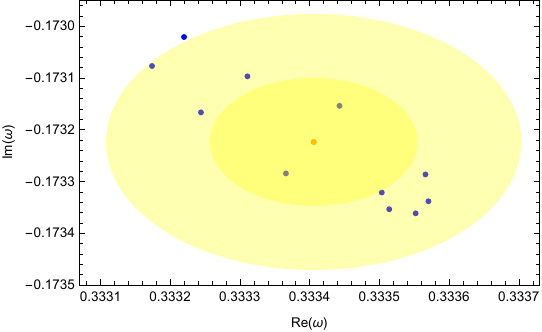}
	\caption{The variance estimate for QNM frequencies with overtone $\hat{\mathcal{n}}=1$. The blue points are the results for the $\hat{\mathcal{n}}=1$ frequencies from the highest 11 iteration orders (the $40$-th order to $50$-th order, here), and the orange point stands for the mean value of them. The light and deep yellow regions indicate $2\sigma$ and $1\sigma$ regions, respectively.}
	\label{fig: AIM_variance}
\end{figure}

Similar to the WKB method, we can use the difference between adjacent iteration order to estimate the precision of the QNM frequencies we get \cite{Konoplya:2019hlu}. Here, we use the variance of the results from the highest iteration orders as the uncertainty of the QNM frequencies. We demonstrate this estimation in Fig. \ref{fig: AIM_variance}. In this plot, we show the mean value and the variance for the $\hat{\mathcal{n}}=1$ QNM frequency under the same parameter choice with Fig. \ref{fig: AIM_order}. The blue points are from the highest $11$ iteration orders, and the orange point is their mean value. The light and deep yellow regions then indicate the $2\sigma$ and $1\sigma$ region, where $\sigma$ is the variance for the real and imaginary parts of the blue points, given by 
\begin{equation}
    \sigma=\sqrt{<\text{re}^2(\omega)>+<\text{im}^2(\omega)>-<\text{re}(\omega)>^2-<\text{im}(\omega)>^2},
    \label{eq: AIM_variance}
\end{equation}
where $<*>$ means the average value of $*$ for the highest $11$ iteration orders.

\begin{figure}[htbp]
	\centering
	\includegraphics[width=0.5\textwidth]{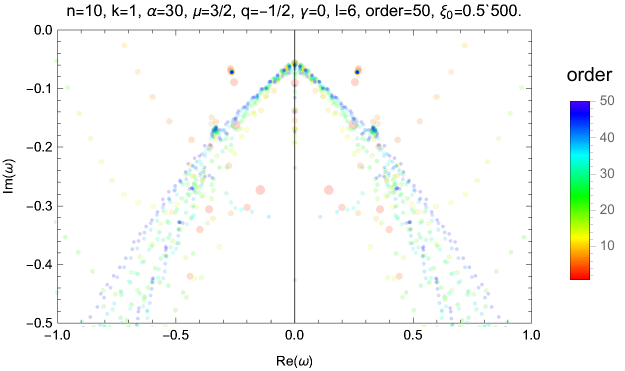}
    \includegraphics[width=0.4\textwidth]{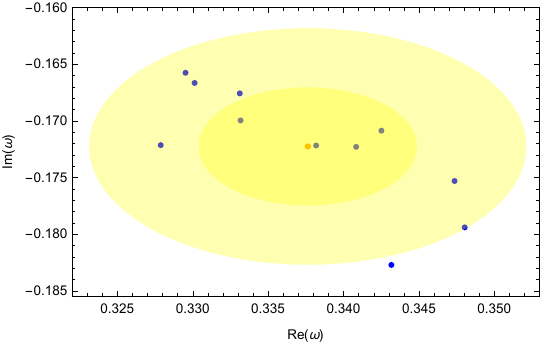}
	\caption{The AIM calculation result with the parameter indicated in the figure. The physical parameters are unchanged, while the expanding positions is chosen to be $\xi_0=0.5$. The left panel shows the numerical result for different orders, and the right panel is for the mean value and variance of the corresponding $\hat{\mathcal{n}}=1$ QNM frequency.}
	\label{fig: AIM_position}
\end{figure}

On the other hand, the precision of AIM heavily depends on the expanding position $\xi_0\in(0,1)$. For a good expanding position, the result can be more accurate, while for a bad expanding position, the precision of the result is much worse. This can be illustrated from Fig. \ref{fig: AIM_order} and Fig. \ref{fig: AIM_position}. As shown in Fig. \ref{fig: AIM_position}, we do the AIM calculation with the same physical parameters as in Fig. \ref{fig: AIM_order}, but expand the equation at $\xi_0=0.5$ instead of $\xi_0=0.4125$. From the left panel, we find that the spot for $\hat{\mathcal{n}}=1$ point is much larger, and we can not read the characteristic frequency for the overtone number $\hat{\mathcal{n}}=2$ from the figure. This is further shown in the right panel, which shows the variance of the highest $11$ iteration orders is more than about $10$ times larger than that of $\xi_0=0.4125$. These suggest that $\xi_0=0.5$ is not a good expanding position. 

\begin{figure}[htbp]
	\centering
	\includegraphics[width=0.88\textwidth]{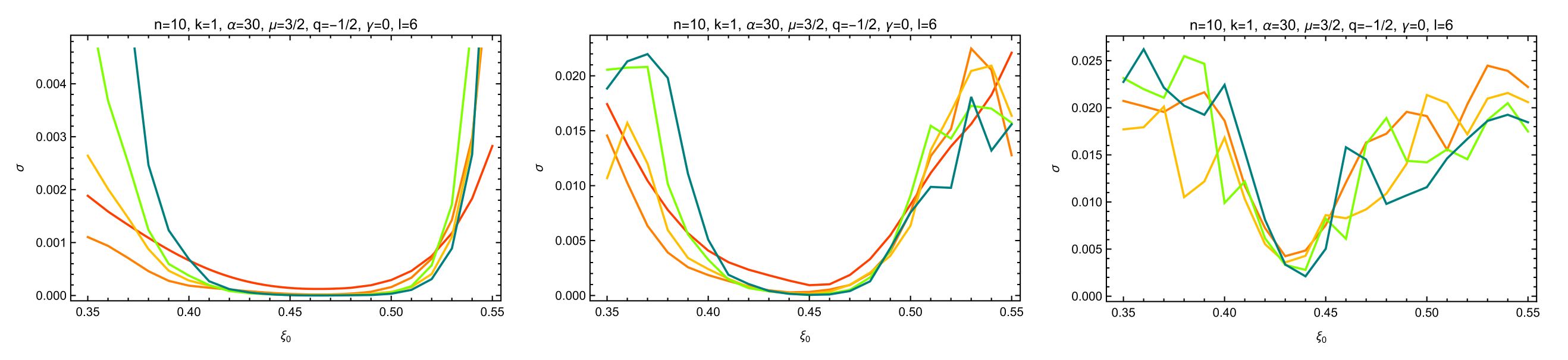}
    \includegraphics[width=0.08\textwidth]{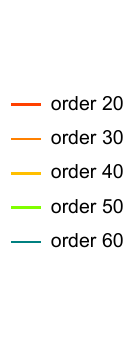}
	\caption{These figures show how the variances defined in \eqref{eq: AIM_variance} of QNM frequencies changes with respect to changing expanding position. The horizontal axis stands for the expanding position, while the vertical axis is for the value of the variance. The three figures are for $\hat{\mathcal{n}}=0$, $\hat{\mathcal{n}}=1$ and $\hat{\mathcal{n}}=2$, respectively. Different color lines stands for different iteration orders.}
	\label{fig: AIM_position_going_through}
\end{figure}

From the analyse above, we see the importance of choosing a proper expanding position to find out the characteristic frequencies. However, up till now, although there are some suggested expanding position for the case where the effective potential has a maxima~\cite{Hakan_Ciftci_2003}, there is no universal method to derive the proper expanding position theoretically for a given AdS-like system. Therefore, in order to overcome this difficulty, we conduct the following analysis. In our calculation, we choose our expanding position by going through the interval $(0,1)$ at a separation of $0.05$ and then choose the one that minimizes the variance mentioned above. The variances for the $\hat{\mathcal{n}}=0$, $\hat{\mathcal{n}}=1$ and $\hat{\mathcal{n}}=2$ QNM frequencies with respect to different expanding position are shown in Fig. \ref{fig: AIM_position_going_through}, respectively. From these plots, we have three observations as follows:
\begin{enumerate}
\item 
    For fixed overtone number $\hat{\mathcal{n}}$, the best expanding position doesn't change markedly with different iteration order.
\item
    The QNM frequency with lower $\hat{\mathcal{n}}$ converges at lower iteration order, and has smaller variance.
\item 
    The best expanding position for different overtone number $\hat{\mathcal{n}}$ is slightly different. However, the choice of $\xi_0$ which works best for higher overtone number $\hat{\mathcal{n}}$ also works fairly well for lower overtone number $\hat{\mathcal{n}}$.

\end{enumerate}

\begin{figure}[htbp]
	\centering
	\includegraphics[width=0.88\textwidth]{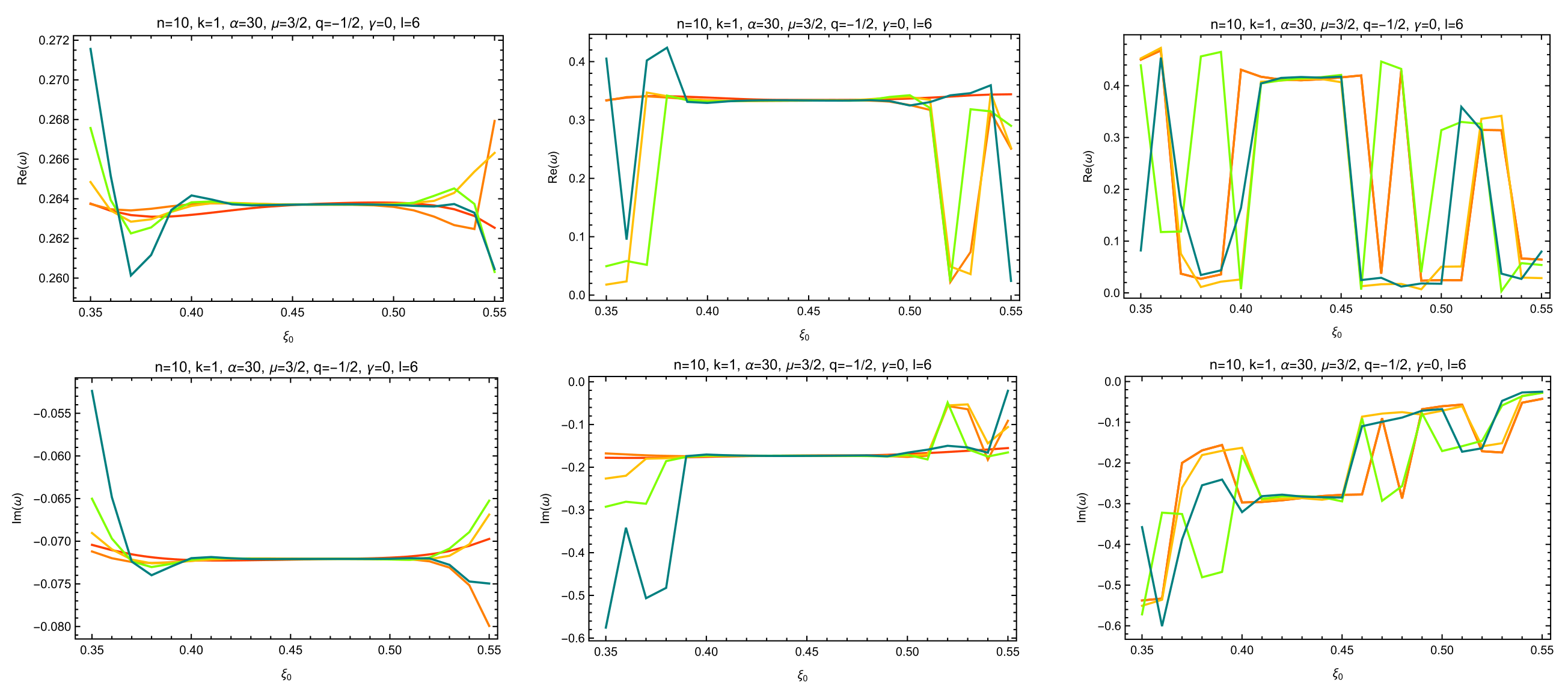}
    \includegraphics[width=0.08\textwidth]{AIM_position_legend.pdf}
	\caption{These figures show how the mean value of QNM frequencies changes with respect to changing expanding position. The horizontal axis stands for the expanding position, while the vertical axis is for the mean value. The six figures are for $\hat{\mathcal{n}}=0$, $\hat{\mathcal{n}}=1$ and $\hat{\mathcal{n}}=2$, from left to right, respectively. The upper panel is for the real part, while the bottom panel is for the imaginary part. Different color lines stands for different iteration orders.}
	\label{fig: AIM_position_mean_going_through}
\end{figure}

We then demonstrate how the expansion position impact the calculation result of QNM frequencies in Fig. \ref{fig: AIM_position_mean_going_through}. These figures shows the mean value of QNM frequencies with respect to different expanding positions. The horizontal axis stands for the expanding position, while the vertical axis stands for the mean value. These six figures are for $\hat{\mathcal{n}}=0$, $\hat{\mathcal{n}}=1$ and $\hat{\mathcal{n}}=2$ from left to right. The upper panel is for the real part, while the bottom panel is for the imaginary part. From these figures, we can see that the expanding position with minima value of variance, the mean value for QNM frequencies doesn't change greatly with respect to expanding position, and the results given by different orders coincide with each other quite well. These relationships between the variance and the mean value for AIM results confirm the variance as a good indicator to choose expanding positions.

For a set of given  physical parameters, in this calculation we want to find out the expanding position that works best for the $\hat{\mathcal{n}}=0$, $\hat{\mathcal{n}}=1$ and $\hat{\mathcal{n}}=2$ QNM frequencies. Based on three observations above, we choose the proper expanding position mainly with the following method. In the first step, we go through all possible $\xi_0$, from $0$ to $1$ under iteration order of $20$, and find out the value of $\xi_0$ that minimizes the variance for the overtone number $\hat{\mathcal{n}}=1$ QNM. This provides a rough estimate of the expanding position and a proper region around the expanding position. In the second step, we use bisection method in this region to find out a more refined expanding position that minimize the variance for the overtone number $\hat{\mathcal{n}}=2$ QNM at iteration order of $30$. It should be noted that for the iteration order of $20$, the variance is averaged over the neighbouring $4$ points, and for the iteration order of $30$, the variance is averaged over the neighbouring $6$ points. After confirming the location of the expansion point, we then use this expanding position for higher order calculation at the iteration order of $50$, and there the variance is averaged over the neighbouring $11$ points.

\section{time-domain analysis}\label{sec: time_domain_analysis}
In this section, we consider the numeric evolution of an initial wave packet in order to investigate the contribution of all modes. We rewrite the wavelike equation (\ref{Schrodinger_equation}) without implying the stationary ansatz ($\Psi\sim e^{-i\omega t}$), i.e., the equation for $\Psi(t,r)$ is given by
\begin{eqnarray}\label{time_equation}
	-\frac{\partial^2\Psi}{\partial t^2}+\frac{\partial^2\Psi}{\partial r_{\star}^2}-V_{\text{eff}}(r)\Psi=0\, ,
\end{eqnarray}
where the effective potential is expressed as (\ref{effective_potential}). The technique of integration of the above wave equation in the time domain was developed by Gaundlach, Price, and Pullin~\cite{Gundlach:1993tp}. In terms of $t$ and $r_{\star}$, we introduce null coordinates $u=t-r_{\star}$ and $v=t+r_{\star}$ so that the black hole horizon $r=r_{+}$ is located at $u=+\infty$. In these coordinates, Eq. (\ref{time_equation}) is written as 
\begin{eqnarray}\label{time_equation_uv}
	-4\frac{\partial^2}{\partial u\partial v}\Psi(u,v)=V_{\text{eff}}(r)\Psi(u,v)\, ,
\end{eqnarray}
where $r$ can be  determined by inverting the relation $r_{\star}(r)=(v-u)/2$, because of the monotonicity of the relation between $r$ and $r_{\star}$.

The two-dimensional wave equation (\ref{time_equation_uv}) can be integrated numerically, using the finite difference method suggested in Refs. \cite{Gundlach:1993tp,Wang:2000dt,Brady:1999wd}. To be specific, Eq. (\ref{time_equation_uv}) can be discretized as
\begin{eqnarray}
	\Psi(N)=\Psi(E)+\Psi(W)-\Psi(S)-h^2V_{\text{eff}}\Big(r\big(\frac{v_N+v_W-u_N-u_E}{4}\big)\Big)\frac{\Psi(E)+\Psi(W)}{8}+\mathcal{O}(h^4)\, ,
\end{eqnarray}
where $S=(u,v)$, $W=(u+h,v)$, $E=(u,v+h)$, $N=(u+h,v+h)$. While the above described integration scheme is efficient for asymptotically flat or de Sitter black holes, for asymptotically AdS black holes like our case, its convergence is too slow~\cite{Konoplya:2011qq}. An alternative integration scheme is put forward in~\cite{Wang:2004bv} which is given by
\begin{eqnarray}\label{integration_scheme}
	\Big[1+\frac{h^2}{16}V_{\text{eff}}(S)\Big]\Psi(N)=\Psi(E)+\Psi(W)-\Psi(S)-\frac{h^2}{16}\Big[V_{\text{eff}}(S)\Psi(S)+V_{\text{eff}}(E)\Psi(E)+V_{\text{eff}}(W)\Psi(W)\Big]+\mathcal{O}(h^4)\, .
\end{eqnarray}
This integration scheme is more stable and in our paper this alternative integration scheme is used. This integration scheme 
can be proved when one uses Taylor expansion at the center of the square. Considering that the behavior of the wave function is not sensitive to the choice of initial data, we set $\Psi(u,v=0)=0$ and use a pulse as an initial perturbation as
\begin{eqnarray}\label{initial_data}
	\Psi(u=0,v)=A\Big(\frac{v-v_1}{v_2-v_1}\Big)^4\Big(1-\frac{v-v_1}{v_2-v_1}\Big)^4\, 
\end{eqnarray}
if $v\in[v_1,v_2]$, and $\Psi(u=0,v)=0$ otherwise. The fourth power is used to ensure that the initial value is smooth at $v_1$ and $v_2$. The symbol $A$ refers to the initial amplitude of the pulse.

\begin{figure}[htbp]
	\centering
	\includegraphics[width=0.7\textwidth]{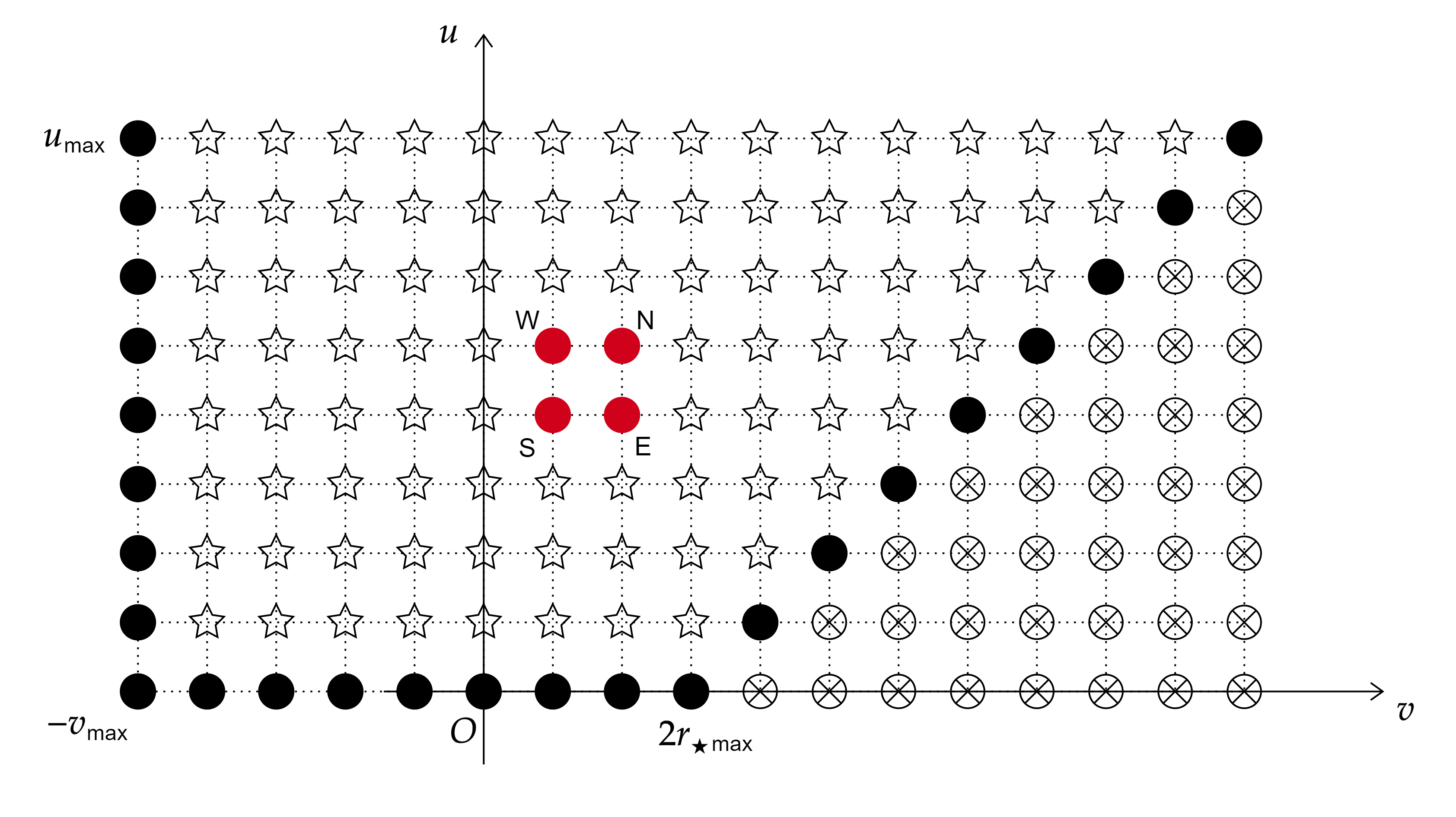}
	\caption{The diagram of numerical grid and the right-angled trapezoidal domain is of our interests. The red dots refer to the east (E), the south (S), the west (W) and the north (N), respectively. From the boundary condition (\ref{boundary_condition}), we have $\Psi=0$ on the sloping waist of the right-angled trapezoidal.}
	\label{grid}
\end{figure}

First, according to the definition of tortoise coordinates, we have
\begin{eqnarray}\label{r_star_max}
	r_{\star}(r)=\int_{r_{\epsilon}}^{r}\frac{\mathrm{d}r^{\prime}}{f(r^{\prime})}\, ,
\end{eqnarray}
where $r_\epsilon$ is chosen as $r_\epsilon=r_{+}+\epsilon$ such that $r_{\star}(r_\epsilon)=0$ and $\epsilon$ is a positive constant that can be given arbitrarily in principle. Hence, the above integral can be worked out numerically although the primitive function of $1/f(r)$ cannot be expressed as an elementary function. It is found that when $r\to+\infty$, $r_{\star}$ tends to a finite constant denoted as $r_{\star\text{max}}$ given by
\begin{eqnarray}\label{r_star_max_max}
	r_{\star\text{max}}=\int_{r_\epsilon}^{+\infty}\frac{\mathrm{d}r^{\prime}}{f(r^{\prime})}\, ,
\end{eqnarray} 
which is determined by $\epsilon$ and $r_\star\to-\infty$ as $r\to r_{+}$. It is worth recalling that our initial condition (\ref{initial_data}) is vanished strictly at $v=2r_{\star\text{max}}$, if $-v_{\text{max}}<v_2<v_1<0$ is selected. Now, we start to build the numerical grid in the Fig. \ref{grid}. In Fig. \ref{grid}, the black spots represent the initial grid points, the stars represent the grid points to be calculated, and the cross product sets represent the forbidden region. Provided that $\epsilon$ is given, we have $N_1$ grid points between the interval $[0,2r_{\star\text{max}}]$, where $r_{\star\text{max}}$ is given by Eq. (\ref{r_star_max_max}), then $h=\mathrm{d}u=\mathrm{d}v=2r_{\star\text{max}}/(N_1-1)$. We have $N_2$ grid points between the interval $[-v_{\text{max}},0]$, where $v_{\text{max}}=(N_2-1)h$. $u_{\text{max}}$ is assumed to be $u_{\text{max}}=(N-1)h=(N_1+N_2-1)h$ for simplicity. 

The evaluation of the potential $V_{\text{eff}}(r)$ is the most challenging part in the computation, which brings more numerical errors. We use the method proposed in~\cite{Gundlach:1993tp,Brady:1999wd} to overcome it. The potential is evaluated at the central radius $r_c$ satisfying
\begin{eqnarray}
	r_{\star}(r_c)=\frac{v_N+v_W-u_N-u_E}{4}=\frac{v_S-u_S}{2}\, .
\end{eqnarray} 
From the Fig. \ref{grid}, it is easy to see that there are $2N-1$ points whose $r$ should be computed in order to get the potential $V_{\text{eff}}(r)$. These points are all on the line segment between the point $(-v_{\text{max}},u_{\text{max}})$ and the point $(2r_{\star\text{max}},0)$. We will number these $2N-1$ points where $(-v_{\text{max}},u_{\text{max}})$ is the first one and $(2r_{\star\text{max}},0)$ is the last one (including the center of the square). Since $r_{\star}(r_{\epsilon})=0$, we use the built-in function $\textbf{FindRoot}$ in \textit{Mathematica} based on $r_{\epsilon}$. After evaluating $r$ along the line segment, we use Eq. (\ref{effective_potential}) to derive the value of $V_{\text{eff}}(r)$ along the line segment and number it in the same order as $r$. Then, the values of the stars in the Fig. \ref{grid} are established as follows. Define $\Psi(u_j,v_k)\equiv\Psi^k_j$, and since $\Psi(u,v=0)=0$, we have
\begin{eqnarray}
	\Psi(u_j,v=0)\equiv\Psi(j,1)\equiv\Psi_j^1=0\, ,
\end{eqnarray}
for $j=1,2,\cdots,N-1,N$. From Eq. (\ref{integration_scheme}), for $k=2,\cdots,N$, we have
\begin{eqnarray}
    \Psi^k_j&=&\Big[1+\frac{h^2}{16}V_{\text{eff}}(k-j+N)\Big]^{-1}\Bigg\{\Psi^k_{j-1}+\Psi^{k-1}_j-\Psi^{k-1}_{j-1}-\frac{h^2}{16}\Big[V_{\text{eff}}(k-j+N)\Psi^{k-1}_{j-1}\nonumber\\
    &&+V_{\text{eff}}(k-j+N+1)\Psi^k_{j-1}+V_{\text{eff}}(k-j+N-1)\Psi^{k-1}_j\Big]\Bigg\}\, ,\quad j=2,\cdots,N\, .
\end{eqnarray}
For $k=N+1,\cdots,2N-1$, we have
\begin{eqnarray}
    \Psi^k_j&=&\Big[1+\frac{h^2}{16}V_{\text{eff}}(k-j+N)\Big]^{-1}\Bigg\{\Psi^k_{j-1}+\Psi^{k-1}_j-\Psi^{k-1}_{j-1}-\frac{h^2}{16}\Big[V_{\text{eff}}(k-j+N)\Psi^{k-1}_{j-1}\nonumber\\
    &&+V_{\text{eff}}(k-j+N+1)\Psi^k_{j-1}+V_{\text{eff}}(k-j+N-1)\Psi^{k-1}_j\Big]\Bigg\}\, ,\quad j=2+k-N,\cdots,N\, .
\end{eqnarray}
There is an issue when the term $V_{\text{eff}}(E)\Psi(E)$ is computed on the grid of sloping waist of the right-angled
trapezoidal. Since $\Psi=0$ on this sloping waist, one can set any value $V_{\text{eff}}$, which does not affect the calculation results. For simplicity, $V_{\text{eff}}(2N-1)=0$ is added into the above numerical scheme. After the integration is completed, the value $\Psi(u_{\text{max}},v)$ is extracted, where $u_{\text{max}}$ is the maximum value of $u$ on the numerical grid. 

\begin{figure}[htbp]
	\centering
	\includegraphics[width=0.5\textwidth]{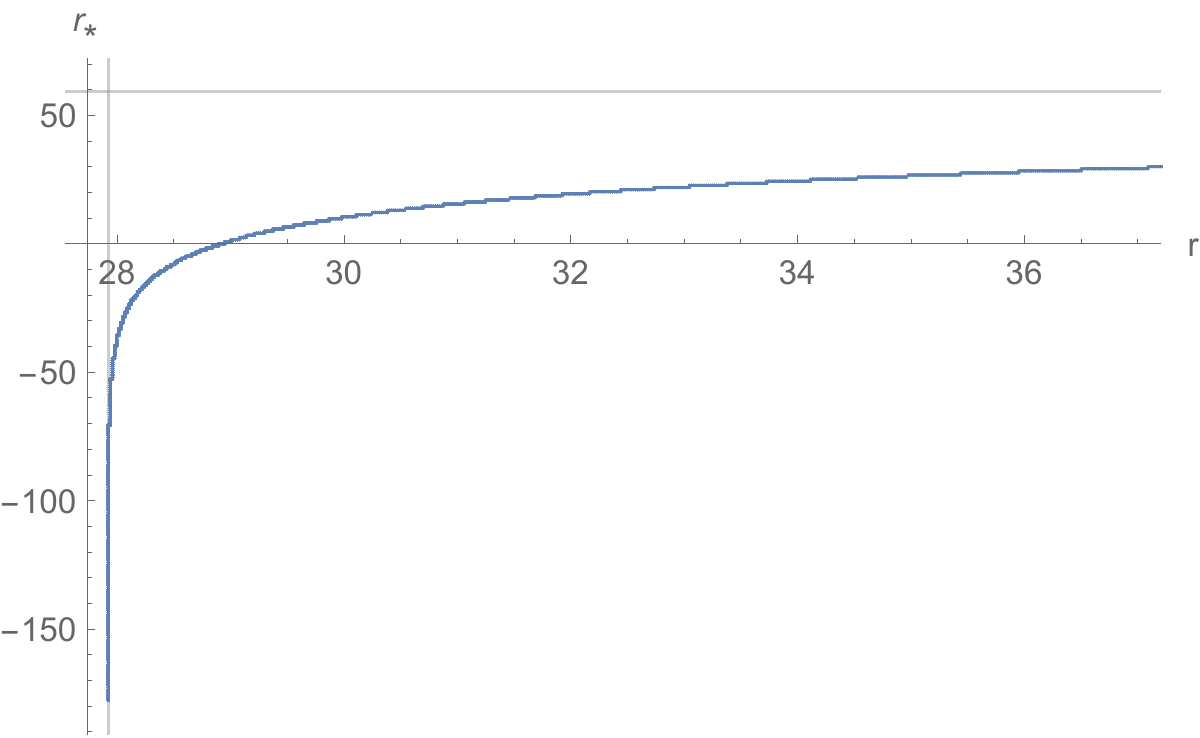}
	\caption{The dependency between $r_{\star}$ and $r$, where the parameters are $n=10$, $\alpha=30$, $\mu=1.5$, $q=-0.5$ and $\epsilon=0.9973$. The two gray lines represent $r_{\star}=r_{\star \text{max}}=59.3221$ and $r=r_{+}=27.9251$, respectively.}
	\label{r_star_r}
\end{figure}

\begin{figure}[htbp]
    \begin{minipage}[t]{0.5\linewidth}
        \centering
        \includegraphics[width=\textwidth]{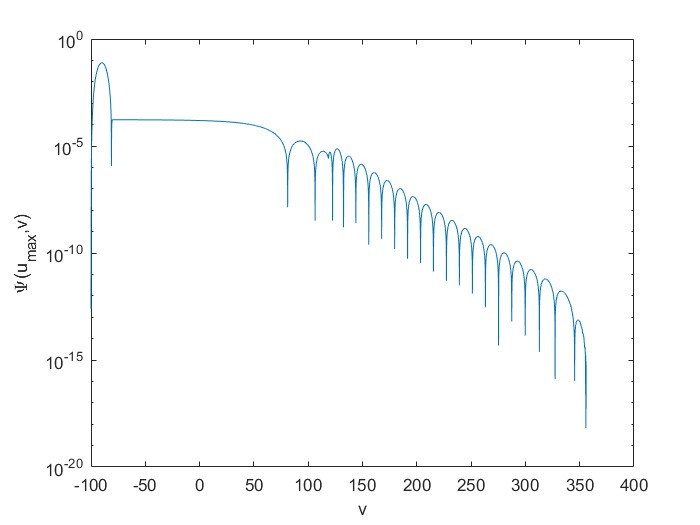}
    \end{minipage}%
    \begin{minipage}[t]{0.5\linewidth}
        \centering
        \includegraphics[width=\textwidth]{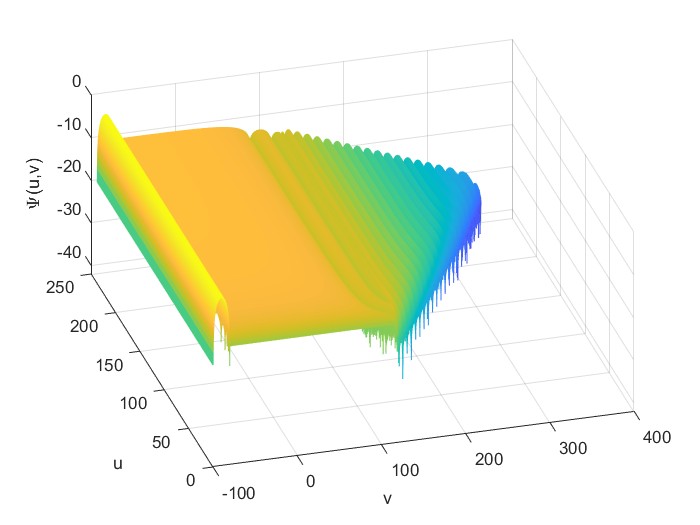}
    \end{minipage}
    \caption{On the left panel, the waveform of $\Psi(u_{\text{max}},v)$ in logarithmic graph is given. The right panel shows the 3D-plot of the waveform of $\Psi(u,v)$ in logarithmic graph. We choose $n=10$, $\alpha=30$, $\mu=1.5$, $q=-0.5$, $\epsilon=0.9973$, $l=6$ and $\gamma=0$. At this case, $A=20$, $v_1=-100$ and $v_2=-80$ are selected as the parameters of the initial wave packet.}
    \label{Psi_umax_v_and_Psi_u_v_3Dplot}
\end{figure}

Now, we give an example to demonstrate how to implement the above algorithm and obtain the corresponding waveform under a specific set of parameters. For the metric function $f(r)$, we choose $n=10$, $\alpha=30$, $\mu=1.5$ and $q=-0.5$. The inner horizon is $r_{-}=1.8991$ and the event horizon is $r_{+}=27.9251$. Choosing $\epsilon=0.9973$, we have $r_{\star\text{max}}=59.3221$. The dependency between $r_{\star}$ and $r$ is shown in Fig. \ref{r_star_r}. Choosing $l=6$ and $\gamma=0$, we get the waveform of $\Psi(u_{\text{max}},v)$ and the 3D-plot of the waveform in Fig. \ref{Psi_umax_v_and_Psi_u_v_3Dplot}. 

\section{characteristic frequency from signal disposition}\label{sec: KT}
To extract the characteristic frequencies from the numerical results in previous section, we implement signal disposition. Following the Ref. \cite{Berti:2007dg}, we present a brief summary of the Prony method and the Kumaresan-Tufts (KT) method for damped sinusoidal signals. Given $N=2p$ equally spaced samples of the signal, with a time sampling interval $T$ between adjacent samples, and indexed by $m=1,2,\cdots N$, we can apply the Prony method to analyze the signal. In the absence of noise, the Prony method assumes that the measured signal $x$ is a linear combination of the ``true" waveform $\Psi$, i.e.,
\begin{eqnarray}\label{signal_equation}
    x[m]=\Psi[m]=\sum_{k=1}^{p}h_kz^{m-1}_k\, ,
\end{eqnarray}
where
\begin{eqnarray}
    h_k&=&A_k e^{i\varphi_k}\, ,\label{h}\\
    z_k&=&e^{(\alpha_k+i\omega_k)T}\, .\label{z}
\end{eqnarray}
The complex parameters $\{h_k,z_k\}$ and the number $p$ of damped sinusoids are to be determined. For $1\le m\le p$ we can rewrite Eq. (\ref{signal_equation}) in matrix form as
\begin{eqnarray}
    \begin{bmatrix}
        z^0_1 & z^0_2 & \cdots & z^0_p\\
        z^1_1 & z^1_2 & \cdots & z^1_p\\
        \vdots & \vdots & \ddots & \vdots\\
        z^{p-1}_1 & z^{p-1}_2 & \cdots&z^{p-1}_p
    \end{bmatrix}\begin{bmatrix}
        h_1\\
        h_2\\
        \vdots\\
        h_p
    \end{bmatrix}=\begin{bmatrix}
        x[1]\\
        x[2]\\
        \vdots\\
        x[p]
    \end{bmatrix}\, .
\end{eqnarray}
In essence, Prony's method is a technique that allows for the determination of the $z_k$'s without requiring nonlinear minimization. Define a polynomial $\mathbf{A}(z)$ of degree $p$ which has the $z_k$'s as its roots:
\begin{eqnarray}\label{polynomial_A}
    \mathbf{A}(z)=\prod_{k=1}^p(z-z_k)\equiv\sum_{\tilde{m}=0}^pa[\tilde{m}]z^{p-\tilde{m}}\, ,
\end{eqnarray}
where $a[0]=1$. It can be shown that the $a[k]$'s are determined from the following matrix equation
\begin{eqnarray}
    \begin{bmatrix}
        x[p] & x[p-1] & \cdots & x[1]\\
        x[p+1] & x[p] & \cdots & x[2]\\
        \vdots & \vdots & \ddots & \vdots\\
        x[2p-1] & x[2p-2] & \dots & x[p]
    \end{bmatrix}\begin{bmatrix}
        a[1]\\
        a[2]\\
        \vdots\\
        a[p]
    \end{bmatrix}
    =-\begin{bmatrix}
        x[p+1]\\
        x[p+2]\\
        \vdots\\
        x[2p]
    \end{bmatrix}\, .
\end{eqnarray}
Then we aim to determine the roots $z_k$ of the polynomial $\mathbf{A}(z)$ [see Eq. (\ref{polynomial_A})]. The damping and frequency are obtained through 
\begin{eqnarray}\label{damping_and_frequency}
    \alpha_k&=&\log|z_k|/T\, ,\nonumber\\
    \omega_k&=&\arctan[\text{Im}(z_k)/\text{Re}(z_k)]/T\, .
\end{eqnarray}
Finally, the amplitudes $A_k$ and phases $\varphi_k$ are found
\begin{eqnarray}\label{amplitudes_and_phases}
    A_k&=&|h_k|\, ,\nonumber\\
    \varphi_k&=&\arctan[\text{Im}(h_k)/\text{Re}(h_k)]\, .
\end{eqnarray}
For most situations, there are more data points than exponential parameters: $N>2p$. One can then use the so called ``least-squares Prony method"~\cite{10.5555/21429} to get the $a[k]$'s from the data and then determine the roots $z_k$, $\alpha_k$, $\omega_k$, $A_k$ and $\varphi_k$ from Eq. (\ref{polynomial_A}), Eq. (\ref{damping_and_frequency}) and Eq. (\ref{amplitudes_and_phases}).

Unfortunately, the results of measurements and numerical simulations will inevitably contain noise. This will make the original and least-squares Prony method no longer applicable. By introducing another characteristic polynomial $\mathbf{B}(z)$, an improved method called KT method is given~\cite{Berti:2007dg,10.5555/21429,article:KT}. The coefficients $a[k]$ of $\mathbf{A}(z)$ are solutions of the forward linear prediction equation given by
\begin{eqnarray}\label{forward_linear_prediction_equation}
    \sum_{\tilde{m}=0}^pa[\tilde{m}]x[n_0-\tilde{m}]=0\, .
\end{eqnarray}
These same exponential waves can be generated in reverse time by the backward linear predictor
\begin{eqnarray}\label{backward_linear_prediction_equation}
    \sum_{\tilde{m}=0}^pb[\tilde{m}]x[n_0-p+\tilde{m}]=0\, ,
\end{eqnarray}
where $b[0]=1$. The characteristic polynomial $\mathbf{B}(z)$ is constructed as 
\begin{eqnarray}
    \mathbf{B}(z)=\sum_{\tilde{m}=0}^pb^{\star}[\tilde{m}]z^{p-\tilde{m}}\, ,
\end{eqnarray}
in which the roots are $z_k=e^{-s^{\star}_k}$ with $s_k=(\alpha_k+i\omega_k)T$ and here $\star$ represents the complex conjugation.

Suppose the measured signal contains additional Gaussian white noise. The noise leads to the deviation of the true zero estimate of the polynomials. As a result, this deviation will cause the real and imaginary parts of characteristic frequency estimates being different from the true ones. By searching for a number of exponential components $L>p$, in which  $p$ represents the actual number of exponential waves in the signal and $L$ is the prediction order of the model, the bias can be significantly reduced in an empirical manner~\cite{Berti:2007dg,10.5555/21429,article:KT}. However, when one uses this process, some extra zeros due to noise will arise. Fortunately, these can be statistically separated by monitoring the zeros of the polynomials $\mathbf{A}(z)$ and $\mathbf{B}(z)$ and the complex conjugate of the reciprocal of these zeros. Singular value decomposition (SVD) can provide the separation. In practice, $p$ in Eq. (\ref{forward_linear_prediction_equation}) and Eq. (\ref{backward_linear_prediction_equation}) is replaced by $L$. We obtain two linear equations with respect to $a[k]$ and $b[k]$, i.e.,
\begin{eqnarray}\label{forward_linear_prediction_equation_L}
    \begin{bmatrix}
        x[L] & x[L-1] & \cdots & x[1]\\
        x[L+1] & x[L] & \cdots & x[2]\\
        \vdots & \vdots & \ddots & \vdots\\
        x[N-1] & x[N-2] & \cdots & x[N-L]
    \end{bmatrix}\begin{bmatrix}
        a[1]\\
        a[2]\\
        \vdots\\
        a[L]
    \end{bmatrix}=-\begin{bmatrix}
        x[L+1]\\
        x[L+2]\\
        \vdots\\
        x[N]
    \end{bmatrix}
\end{eqnarray}
and
\begin{eqnarray}\label{backward_linear_prediction_equation_L}
    \begin{bmatrix}
        x[2] & x[3] & \cdots & x[L+1]\\
        x[3] & x[4] & \cdots & x[L+2]\\
        \vdots & \vdots & \ddots & \vdots\\
        x[N-L+1] & x[N-L+2] & \cdots & x[N]
    \end{bmatrix}\begin{bmatrix}
        b[1]\\
        b[2]\\
        \vdots\\
        b[L]
    \end{bmatrix}=-\begin{bmatrix}
        x[1]\\
        x[2]\\
        \vdots\\
        x[N-L]
    \end{bmatrix}\, ,
\end{eqnarray}
where $n_0$ is given by $L+1,L+2,\cdots,N-1,N$ in Eq. (\ref{forward_linear_prediction_equation}) and Eq. (\ref{backward_linear_prediction_equation}). We express $\mathbf{X}$ which is the coefficient matrix of Eq. (\ref{forward_linear_prediction_equation_L}) or Eq. (\ref{backward_linear_prediction_equation_L}) as
\begin{eqnarray}
    \mathbf{X}=\mathbf{U}\mathbf{S}\mathbf{V}^H\, ,
\end{eqnarray}
where $\mathbf{U}$ is a $(N-L)\times(N-L)$ dimensional matrix, $\mathbf{S}$ is a $(N-L)\times L$ dimensional matrix and $\mathbf{V}$ is a $L\times L$ dimensional matrix with superscript $H$ stands for the Hermitian conjugation. The singular values on the diagonal $(s_1,\cdots,s_p,s_{p+1},\cdots,s_L)$ are arranged in decreasing order. Noise will be reduced by considering the reduced rank approximation
\begin{eqnarray}
    \hat{\mathbf{X}}=\mathbf{U}\hat{\mathbf{S}}\mathbf{V}^H
\end{eqnarray}
with
\begin{eqnarray}
    \hat{\mathbf{S}}=\begin{bmatrix}
        \hat{\mathbf{S}}_p & \mathbf{0}\\
        \mathbf{0} & \mathbf{0}
    \end{bmatrix}_{(N-L)\times L}\, ,
\end{eqnarray}
where $\hat{\mathbf{S}}_p$ is the top-left $p\times p$ of $\mathbf{S}$. A better estimate for the coefficients $a[k]$ and $b[k]$ is then
\begin{eqnarray}\label{Moore_Penrose_inverse}
    \hat{\mathbf{a}}=-\hat{\mathbf{X}}^{+}\mathbf{x}\, ,
\end{eqnarray}
where $\hat{\mathbf{X}}^{+}$ is the Moore-Penrose inverse of $\hat{\mathbf{X}}$ and $\hat{\mathbf{a}}$ stands for $a[k]$'s or $b[k]$'s. This is the basic idea for the Kumaresan-Tufts methods~\cite{Berti:2007dg,10.5555/21429,article:KT}. It should be mentioned that SVD and Moore-Penrose inverse are both built-in functions in \textit{Matlab}, writing as \textbf{svd} and \textbf{pinv}. As for the KT method in practice, we choose $L=N/3$ in order to minimize the variance~\cite{Berti:2007dg} where $N$ is the number of samples.

\begin{figure}[htbp]
	\centering
	\includegraphics[width=0.5\textwidth]{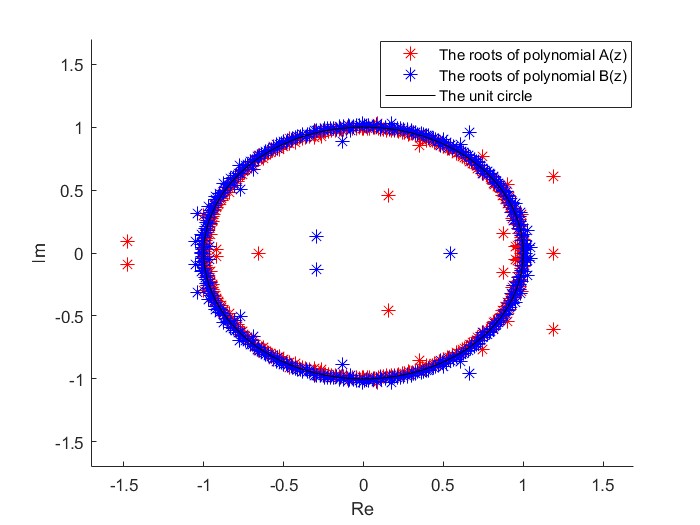}
	\caption{The red asterisks stand for the roots of polynomial $\mathbf{A}(z)$. The blue asterisks stand for the roots of polynomial $\mathbf{B}(z)$.}
	\label{roots_polynomial_A_and_B_old}
\end{figure}

Now, taking the parameters given in Fig. \ref{Psi_umax_v_and_Psi_u_v_3Dplot} as an example, we will use the KT method to extract quasi-normal frequencies. First, we select an appropriate sampling time interval by observing the data in the left panel of Fig. \ref{Psi_umax_v_and_Psi_u_v_3Dplot}. For the entire interval of $v$, we choose the line segments proportionally as $[0.53,0.88]$ to extract the characteristic frequencies where $0.53$ means that we start to extract at $v_{\text{initial}}=132.8815$ and $0.88$ means that we end with $v_{\text{final}}=298.9833$. It is noted that $v_{\text{initial}}=0.53\times(v_{\text{end}}-v_{\text{start}})+v_{\text{start}}$ and $v_{\text{final}}=0.88\times(v_{\text{end}}-v_{\text{start}})+v_{\text{start}}$ in which $v_{\text{start}}$ is the time when the numerical simulation starts, $v_{\text{end}}$ is the time when the numerical simulation ends. It is observed that the ringdown begins at around $v_{\text{initial}}=132.8815$. For this example, $N=1200$ is the number of samples and the prediction order is $L=400$ as previously mentioned. We obtain $a[k]$'s and $b[k]$'s from Eq. (\ref{forward_linear_prediction_equation_L}) and Eq. (\ref{backward_linear_prediction_equation_L}) so the roots of polynomial $\mathbf{A}(z)$ and $\mathbf{B}(z)$ can be found. These roots are irregularly distributed on both sides of the unit circumference which are shown in Fig. \ref{roots_polynomial_A_and_B_old} . At this time, $a[k]$'s and $b[k]$'s are not modified by the SVD method. 

\begin{figure}[htbp]
    \begin{minipage}[t]{0.5\linewidth}
        \centering
        \includegraphics[width=\textwidth]{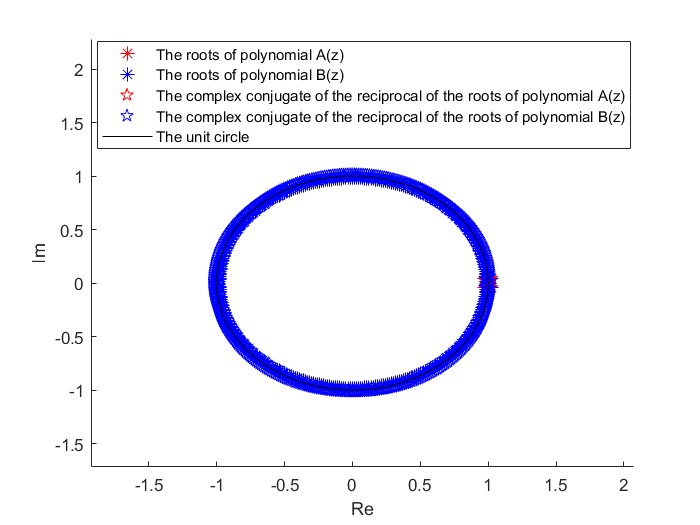}
    \end{minipage}%
    \begin{minipage}[t]{0.5\linewidth}
        \centering
        \includegraphics[width=\textwidth]{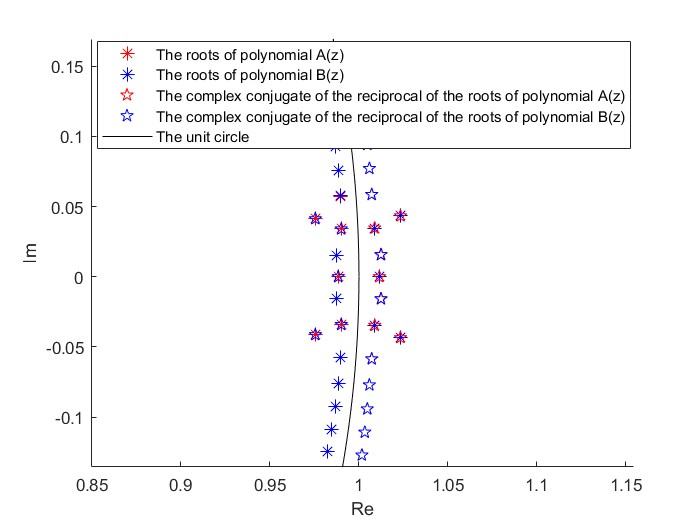}
    \end{minipage}
    \caption{The new roots are shown after the SVD decomposition. The red asterisks stand for the roots of polynomial $\mathbf{A}(z)$. The blue asterisks stand for the roots of polynomial $\mathbf{B}(z)$. The red five-pointed stars stand for the complex conjugate of the reciprocal of the roots of polynomial $\mathbf{A}(z)$. The blue five-pointed stars stand for the complex conjugate of the reciprocal of the roots of polynomial $\mathbf{B}(z)$. The image on the right is an enlargement of the image on the left near the point $(1,0)$.}
    \label{roots_polynomial_A_and_B}
\end{figure}

Then, choosing physical number of exponential waves as $p=5$ and using the SVD method, we will have the new $a[k]$'s and $b[k]$'s [see Eq. (\ref{Moore_Penrose_inverse})] and then derive the new roots of polynomial $\mathbf{A}(z)$ and $\mathbf{B}(z)$ which are shown in Fig. \ref{roots_polynomial_A_and_B}. To find the physical frequency, we also need to find the conjugate roots of these roots which are also shown in Fig. \ref{roots_polynomial_A_and_B}. After applying the SVD method, it is found that almost all of the red asterisks are coincident with blue asterisks except five red asterisks which are coincident with blue five-pointed stars. Just as explained previously, these five points correspond to physical frequencies while other points are not physical which are considered to be the noise. The theoretical support for the above statement is that for both polynomials $\mathbf{A}(z)$ and $\mathbf{B}(z)$, zeros due to the noise tend to stay within the unit circle, whereas the true zeros due to the exponential signal form complex conjugate pairs inside and outside the unit circle. This is in general as a result of the fact that the statistics of a stationary random process do not change under time reversal~\cite{Berti:2007dg}. 

\begin{figure}[htbp]
	\centering
	\includegraphics[width=0.5\textwidth]{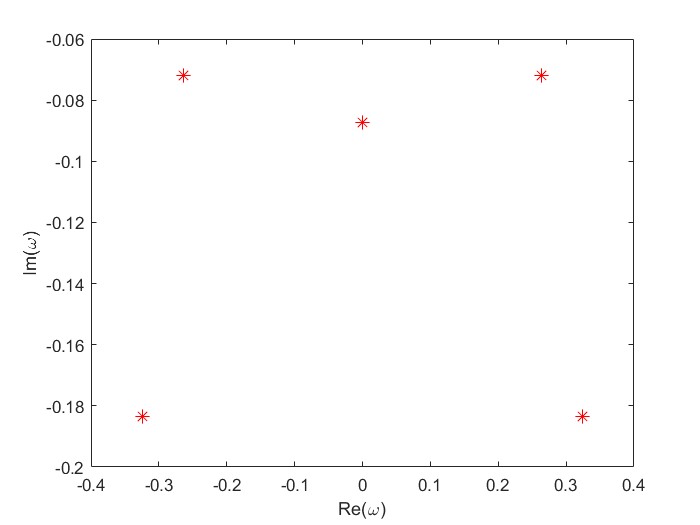}
	\caption{These five characteristic frequencies are symmetrical in terms of the imaginary axis. The value of these characteristic frequencies are $-0.0873i$, $\pm0.2635-0.07208i$ and $\pm0.3248-0.1836i$, respectively.}
	\label{characteristic_frequency_KT}
\end{figure}

\begin{figure}[htbp]
	\centering
	\includegraphics[width=0.5\textwidth]{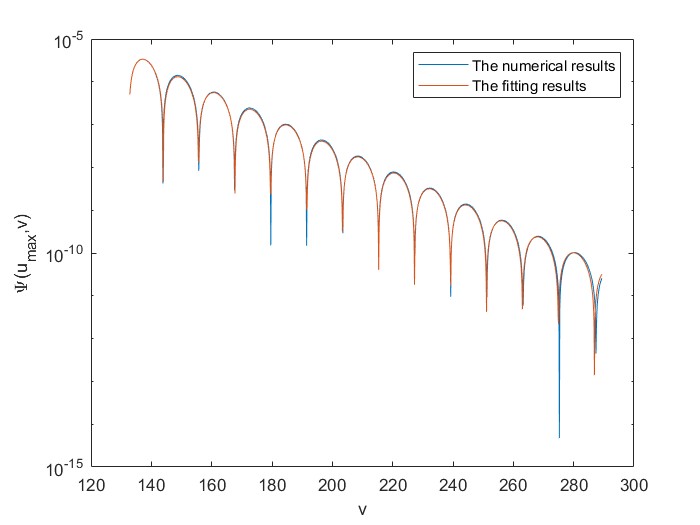}
	\caption{The comparation among the results of numerical calculations and the results of fitting. The blue curve stands for the numerical results and the orange curve stands for the fitting results.}
	\label{comparation_numerical_and_fitting}
\end{figure}

Following such a process, we have five physical frequencies in all. Substituting them into Eq. (\ref{signal_equation}), $h_k$ will be acquired with $x[i]$ coming from our collected samples. Finally, we have the damping and the frequency through Eq. (\ref{damping_and_frequency}) where the time sampling interval $T=0.1305$ in this example. The physical frequencies are displayed in Fig. \ref{characteristic_frequency_KT}. Furthermore, the amplitudes $A_k$ and the phases $\varphi_k$ are determined from Eqs. (\ref{amplitudes_and_phases}). The comparation among the results of numerical calculations and the results of fitting are represented in Fig. \ref{comparation_numerical_and_fitting}. The model has a good fit and we find that the physical frequencies derived in this section is compatible with the AIM. This will be explained in more detail in Sec. \ref{sec: numerical_result_and_analysis}.

Last but not least, determining the beginning of quasi-normal ringing is somewhat ambiguous since the quasi-normal stage is the one that can not be defined exactly on the one hand. On the other hand, higher overtones damp quickly and are exponentially suppressed. As a result, they are difficult to distinguish from numerical errors within the fitting approach, making them challenging to identify~\cite{Zhidenko:2009zx}. In fact, we are able to calculate only two or, sometimes, three longest-living frequencies including the point on the imaginary axis. However, through our practice, we find that using KT method is suitable to find more characteristic frequencies than the least-squares Prony method in our model.

\section{numerical result and analysis}\label{sec: numerical_result_and_analysis}
In this section, we will show some typical results. In order to  provide how a single physical parameter affects the characteristic frequencies, we provide benchmark parameters as spacetime dimension $n=10$, coupling constant $\alpha=30$, black hole mass parameter $\mu=1.5$, black hole charge parameter $q=-0.5$, and ``quantum numbers" $l=6$ and $\gamma=0$ which are used in Sec. \ref{sec: AIM} and Sec. \ref{sec: time_domain_analysis}. It should be noted that there is nothing special about this set of benchmark parameters.

\begin{figure}[htbp]
	\centering
	\includegraphics[width=0.33\textwidth]{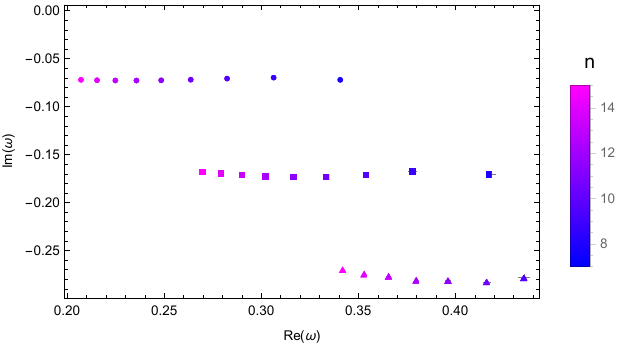}
    \includegraphics[width=0.33\textwidth]{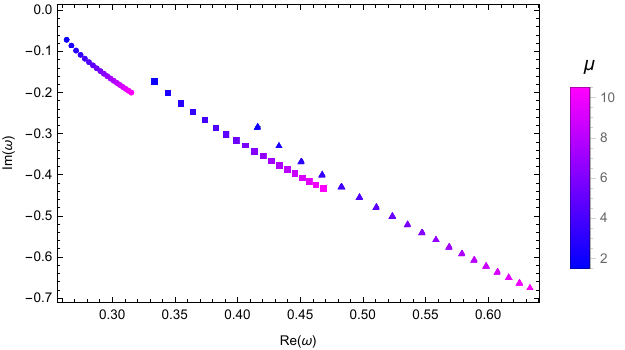}
    \includegraphics[width=0.33\textwidth]{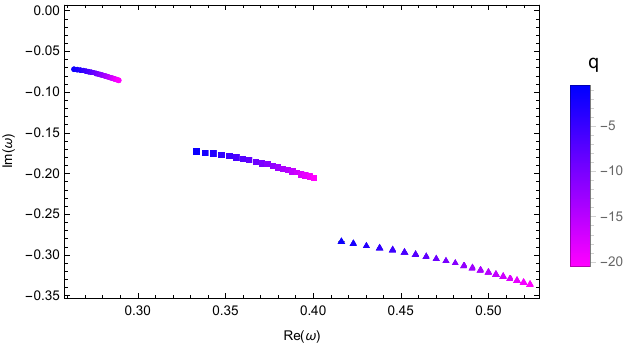}
    \includegraphics[width=0.33\textwidth]{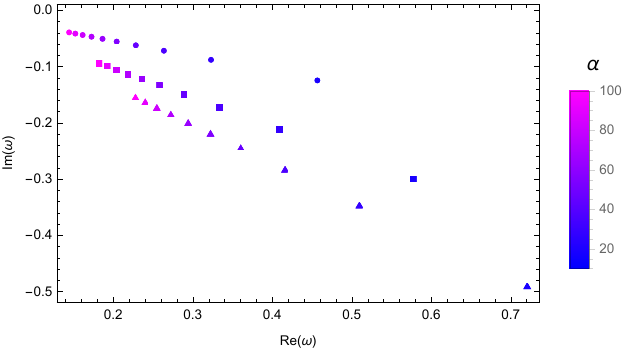}
    \includegraphics[width=0.33\textwidth]{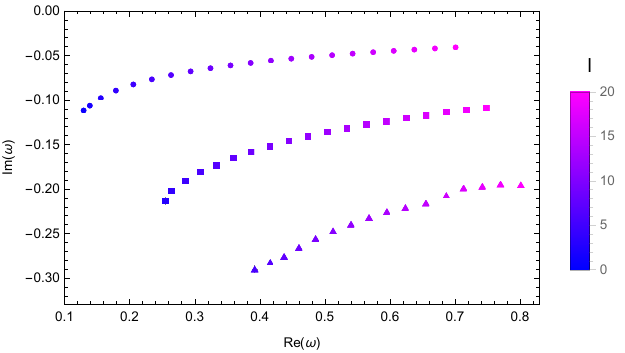}
    \includegraphics[width=0.33\textwidth]{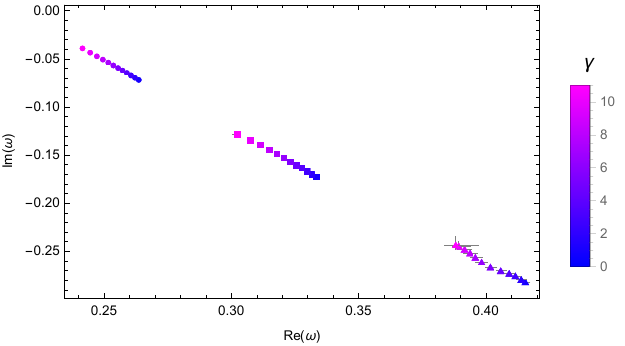}
	\caption{QNM frequencies of overtone number $\hat{\mathcal{n}}=0, 1, 2$ with different physical parameters. The six figures are for the spacetime dimension $n$, the mass $\mu$, the charge $q$, the Gauss-Bonnet coupling constant $\alpha$, and the ``quantum numbers" $l$ and $\gamma$. Results with overtone number $\hat{\mathcal{n}}=0, 1, 2$ are marked by circle, rectangle and triangle, repesctively.}
	\label{fig: AIM_data_graph}
\end{figure}

In Fig. \ref{fig: AIM_data_graph}, we show the first three order characteristic frequencies obtained from AIM excepting the pure imaginary modes where we show the pure imaginary modes in the tables (see Tab. \ref{tab: AIM_n}-Tab. \ref{tab: AIM_gamma}). These six figures describe the relationship between  characteristic  frequencies and the six parameters $n$, $\mu$, $q$, $\alpha$, $l$ and $\gamma$, respectively. The horizontal axis in the figure represents the real part of the frequency, while the vertical axis represents the imaginary part of the frequency. The markers of circle, rectangle and triangle stand for overtone numbers $\hat{\mathcal{n}}=0$, $\hat{\mathcal{n}}=1$ and $\hat{\mathcal{n}}=2$, respectively.

At first glance, it is found that the relationship between frequencies and parameters roughly shows a linear relationship for fixed overtones (The curvature of the curve is low.). Since all the imaginary part of frequencies we get are negative, for  convenience of discussion, in later discussions the imaginary part of frequencies refers to the absolute value of the imaginary part without any special instructions.

For the dimension of spacetime $n$, we find that the imaginary part of frequencies are not significantly dependent on the dimension of  spacetime while the real part decreases as the dimension of spacetime increases. So we can say the lifetime of QNMs are not remarkably dependent on the number of extra dimension of spacetime. For the black hole mass parameter $\mu$, we find that the imaginary part of frequencies increases with the increase of the mass, while the real part increases with the increase of the mass. Therefore, the lifetime of QNMs decreases with the increase of the mass. For the parameter of black hole charge $q$, we see a weak dependence on the imaginary part of frequencies for the charge. The imaginary part of frequencies increases with the increase of the charge $|q|$, while the real part increases with the increase of the charge. The lifetime of QNMs decreases with the increase of the charge $|q|$. For the Gauss-Bonnet coupling constant $\alpha$, it is found that $\sqrt{\alpha}\omega$ remain constants when only $\alpha$ varies, as we stated before. The lifetime of QNMs increases with the increase of the Gauss-Bonnet coupling constant. Additionally, the Gauss-Bonnet coupling constant has more impacts on the imaginary part of the third overtone than the first and the second overtones. The above studies are showed to illustrate the relationship between the characteristic frequencies and the four physical parameters $n$, $\mu$, $q$ and $\alpha$ where they are all appear at the metric function $f(r)$.

There are two parameters $l$ and $\gamma$ that can effect the frequencies which are called the ``quantum numbers". For the parameter $l$, we find that the imaginary part of frequencies decreases with the increase of $l$, while the real part increases with the increase of $l$. The different performance is that the linear relationship between the imaginary part and the real part disappears especially for $l\to0$. The lifetime of QNMs increases with the increase of $l$. For another quantum parameter $\gamma$, we find that the imaginary part of frequencies decreases with the increase of $\gamma$, while the real part decreases with the increase of $\gamma$. Interestingly, it is found that the slopes of the three lines of three overtones are almost the same within the range of errors. So, it can be said that rain and dew are evenly distributed on the each overtone for the quantum parameter $\gamma$ while the Gauss-Bonnet coupling constant is not.

As for those pure imaginary modes, we place the first pure imaginary frequency on the right side of the table and these modes do not participate in the sorting of overtones. Here, we provide some interesting discoveries among the results. For the dimension $n$, it can be seen that the relationship between frequency and $n$ is not monotonic. For the mass $\mu$, we notice that as the mass increases, the fundamental mode changes from the non-imaginary axis mode dominating to the imaginary axis mode dominating. For the ``quantum number" $l$, it is found that when $l\ge 7$, the pure imaginary frequency is dissipated, which is confirmed at the side of numerical integration method.

Considering the limitation of the numerical integration method, in our analysis AIM is the main approach to derive the QNM frequencies and the numerical integration method is an auxiliary method. In other words, we use the numerical integration method to check the consistence of the frequencies obtained by two methods and calculate the corresponding errors. The relative error formula used in this paper is 
\begin{eqnarray}\label{relative_error}
    \delta=\frac{|\omega_{\text{AIM}}-\omega_{\text{NIM}}|}{|\omega_{\text{AIM}}|}\times100\%\, ,
\end{eqnarray}
where AIM refers to the asymptotic iteration method and NIM refers to the numerical integration method. A typical example is that the spacetime dimension $n=10$, coupling constant $\alpha=30$, black hole mass parameter $\mu=1.5$, black hole charge parameter $q=-0.5$, and ``quantum numbers" $l=6$ and $\gamma=0$. Using Eq. (\ref{relative_error}), we get the relative error $\delta_0=0.072\%$, $\delta_1=3.6\%$ and $\delta_{\text{pure}}=1.9\%$ where $0$ and $1$ refer to the overtone number and ``pure" refers to the pure imaginary mode. By applying the relative error formula (\ref{relative_error}), we observe that the error is within an acceptable range. The corresponding calculation results are not shown but they are all small especially for the overtone number $\hat{\mathcal{n}}=0$. 

It has been shown that one can obtain a four dimensional wave-like equation on the manifold $M^4$ by using the characteristic tensors in extra dimensions [see Eq. (\ref{separate_variables}) and Eq. (\ref{gamma})]. In Appendix \ref{KG}, we study a toy model (Klein-Gordon equation) to compute the QNM frequencies in four dimensional spacetime so as to compare our model with the test model. The Klein-Gordon toy model is a naive calculation of the perturbation propagating on the $M^4$ ignoring the dynamics on the compactification part of the spacetime. The results are shown in Fig. \ref{fig: AIM_KG_data_graph} with the same parameters excepting the ``quantum number" $\gamma$, since the ``quantum number" $\gamma$ does not appear in the KG model. 

Comparing the various figures, we found the following differences: For the dimension of the spacetime $n$, it is found that the lifetimes of the Klein-Gorden test field for the low dimensional spacetime are more dependent on the dimension of the spacetime than the tensor perturbation model. For the parameter of mass $\mu$, the performance among these two models are very different. The biggest difference is that the overtone $\hat{\mathcal{n}}=1$ of the Klein-Gorden test field has the non monotonous behavior. The shapes of the three lines are completely different for the Klein-Gorden test field. But for the tensor perturbation model, the corresponding three lines exhibit same patterns.  For the parameter of charge $q$, the three lines show a similar shape in both models. However, the same parameter ranges make the characteristic frequency variation ranges larger in the Klein-Gorden model. For the Gauss-Bonnet coupling constant $\alpha$, we think the behaviors are the same of the tensor perturbation model and the Klein-Gorden model. For the angular quantum number $l$, the Klein-Gorden model has an irregular behavior for low number $l$ but the tensor perturbation model doesn't.

So we see that the tensor perturbation model give significantly different results compared with the toy Klein-Gordon model, verifying the influence of the dynamics of the compactification part on the $4$-dimensional part $M^4$. If we can find any possibility of observing this signal, then the corresponding QNM frequency might be a probe to distinguish the difference between the full tensor perturbation and the toy Klein-Gordon equation, and thus be a method to find the signal of extra dimension.

\begin{table*}
    \caption{QNM frequencies for the dimension $n$. The results are calculated with asymptotic iteration method of $50$ iteration order. "-" symbol indicates that AIM can't predict the result with enough precision with corresponding parameter choice.}
    \begin{ruledtabular}
    \begin{tabular}{ccccccccc}
    	\label{tab: AIM_n}
         &\multicolumn{2}{c}{$\hat{\mathcal{n}}=0$}&\multicolumn{2}{c}{$\hat{\mathcal{n}}=1$}&\multicolumn{2}{c}{$\hat{\mathcal{n}}=2$}&\multicolumn{2}{c}{first pure imaginary}\\
         $n$&$\omega$&$\sqrt{\sigma_{re}^2+\sigma_{im}^2}$&$\omega$
        &$\sqrt{\sigma_{re}^2+\sigma_{im}^2}$&$\omega$
        &$\sqrt{\sigma_{re}^2+\sigma_{im}^2}$&$\omega$
        &$\sqrt{\sigma_{re}^2+\sigma_{im}^2}$\\ 
    \hline
         7 & {0.3407 - 0.07223 i} & {3.436e-4} & {0.4174 - 0.1707 i} & {3.758e-3} & - & - &  {- 0.06898 i} & {5.593e-4} \\
         8 & {0.3064 - 0.06988 i} & {3.030e-4} & {0.3779 - 0.1675 i} & {2.716e-3} & - & - &  {- 0.08114 i} & {4.715e-4} \\
         9 & {0.2824 - 0.07091 i} & {5.328e-5} & {0.3538 - 0.1715 i} & {4.615e-4} & {0.4353 - 0.2786 i} & {4.363e-3} &  {- 0.08474 i} & {1.296e-3} \\
         10 & {0.2637 - 0.07208 i} & {2.256e-5} & {0.3336 - 0.1733 i} & {1.937e-4} & {0.4161 - 0.2830 i} & {2.627e-3} &  {- 0.08566 i} & {7.629e-4} \\
         11 & {0.2485 - 0.07270 i} & {2.654e-5} & {0.3167 - 0.1734 i} & {2.333e-4} & {0.3962 - 0.2817 i} & {2.081e-3} &  {- 0.08465 i} & {1.058e-4} \\
         12 & {0.2357 - 0.07293 i} & {1.401e-5} & {0.3023 - 0.1726 i} & {1.278e-4} & {0.3798 - 0.2812 i} & {1.295e-3} &  {- 0.08328 i} & {7.548e-5} \\
         13 & {0.2249 - 0.07286 i} & {1.997e-5} & {0.2900 - 0.1714 i} & {1.818e-4} & {0.3656 - 0.2772 i} & {1.563e-3} &  {- 0.08167 i} & {1.167e-5} \\
         14 & {0.2154 - 0.07261 i} & {1.247e-5} & {0.2792 - 0.1698 i} & {1.177e-4} & {0.3530 - 0.2748 i} & {1.002e-3} &  {- 0.08002 i} & {8.808e-6} \\
         15 & {0.2072 - 0.07223 i} & {1.990e-5} & {0.2697 - 0.1681 i} & {1.852e-4} & {0.3419 - 0.2704 i} & {1.589e-3} &  {- 0.07838 i} & {1.528e-6} \\
    \end{tabular}
    \end{ruledtabular}
\end{table*}

\begin{table*}
    \label{tab: AIM_mu}
    \caption{QNM frequencies for the mass $\mu$. The results are calculated with asymptotic iteration method of $50$ iteration order.}
    \begin{ruledtabular}
    \begin{tabular}{ccccccccc}
         &\multicolumn{2}{c}{$\hat{\mathcal{n}}=0$}&\multicolumn{2}{c}{$\hat{\mathcal{n}}=1$}&\multicolumn{2}{c}{$\hat{\mathcal{n}}=2$}&\multicolumn{2}{c}{first pure imaginary}\\
         $\mu$&$\omega$&$\sqrt{\sigma_{re}^2+\sigma_{im}^2}$&$\omega$
        &$\sqrt{\sigma_{re}^2+\sigma_{im}^2}$&$\omega$
        &$\sqrt{\sigma_{re}^2+\sigma_{im}^2}$&$\omega$
        &$\sqrt{\sigma_{re}^2+\sigma_{im}^2}$\\ 
    \hline
         1.5 & {0.2637 - 0.07208 i} & {2.256e-5} & {0.3336 - 0.1733 i} & {1.937e-4} & {0.4161 - 0.283 i} & {2.627e-3} &  {- 0.08566 i} & {7.629e-4} \\
         2. & {0.2675 - 0.08634 i} & {5.619e-6} & {0.3444 - 0.2027 i} & {4.915e-5} & {0.4331 - 0.3297 i} & {4.489e-4} &  {- 0.09916 i} & {2.898e-6} \\
         2.5 & {0.2712 - 0.0983 i} & {1.978e-6} & {0.3549 - 0.2274 i} & {2.017e-5} & {0.4510 - 0.3667 i} & {1.806e-4} &  {- 0.1095 i} & {1.451e-7} \\
         3. & {0.2748 - 0.1087 i} & {1.440e-6} & {0.3647 - 0.2490 i} & {1.427e-5} & {0.4676 - 0.3992 i} & {1.221e-4} &  {- 0.1177 i} & {1.477e-8} \\
         3.5 & {0.2782 - 0.1180 i} & {5.341e-7} & {0.3740 - 0.2681 i} & {5.295e-6} & {0.4830 - 0.4280 i} & {4.907e-5} &  {- 0.1244 i} & {7.514e-9} \\
         4. & {0.2815 - 0.1265 i} & {6.627e-7} & {0.3827 - 0.2855 i} & {6.665e-6} & {0.4973 - 0.4540 i} & {6.095e-5} &  {- 0.1301 i} & {1.025e-9} \\
         4.5 & {0.2847 - 0.1343 i} & {3.598e-7} & {0.3910 - 0.3014 i} & {3.943e-6} & {0.5108 - 0.4778 i} & {3.637e-5} &  {- 0.1351 i} & {6.371e-10} \\
         5. & {0.2877 - 0.1415 i} & {2.322e-7} & {0.3989 - 0.3161 i} & {2.638e-6} & {0.5236 - 0.4997 i} & {2.338e-5} &  {- 0.1395 i} & {3.905e-10} \\
         5.5 & {0.2906 - 0.1482 i} & {3.925e-7} & {0.4064 - 0.3299 i} & {4.425e-6} & {0.5357 - 0.5202 i} & {4.033e-5} &  {- 0.1434 i} & {1.511e-10} \\
         6. & {0.2934 - 0.1546 i} & {2.911e-7} & {0.4136 - 0.3427 i} & {3.342e-6} & {0.5473 - 0.5394 i} & {2.977e-5} &  {- 0.1470 i} & {1.201e-10} \\
         6.5 & {0.2961 - 0.1606 i} & {2.283e-7} & {0.4205 - 0.3549 i} & {2.612e-6} & {0.5583 - 0.5574 i} & {2.277e-5} &  {- 0.1503 i} & {9.093e-11} \\
         7. & {0.2988 - 0.1664 i} & {1.848e-7} & {0.4272 - 0.3664 i} & {2.094e-6} & {0.5689 - 0.5745 i} & {1.797e-5} &  {- 0.1534 i} & {6.594e-11} \\
         7.5 & {0.3013 - 0.1718 i} & {1.528e-7} & {0.4336 - 0.3774 i} & {1.714e-6} & {0.5791 - 0.5908 i} & {1.459e-5} &  {- 0.1562 i} & {4.540e-11} \\
         8. & {0.3038 - 0.1771 i} & {1.284e-7} & {0.4398 - 0.3878 i} & {1.431e-6} & {0.5889 - 0.6063 i} & {1.213e-5} &  {- 0.1589 i} & {2.908e-11} \\
         8.5 & {0.3062 - 0.1821 i} & {2.611e-7} & {0.4458 - 0.3978 i} & {3.013e-6} & {0.5984 - 0.6211 i} & {2.628e-5} &  {- 0.1614 i} & {2.255e-11} \\
         9. & {0.3086 - 0.1869 i} & {2.294e-7} & {0.4517 - 0.4074 i} & {2.628e-6} & {0.6075 - 0.6353 i} & {2.275e-5} &  {- 0.1638 i} & {4.043e-12} \\
         9.5 & {0.3109 - 0.1916 i} & {2.036e-7} & {0.4573 - 0.4167 i} & {2.318e-6} & {0.6164 - 0.6490 i} & {1.997e-5} &  {- 0.1661 i} & {1.330e-11} \\
         10. & {0.3131 - 0.1961 i} & {1.823e-7} & {0.4629 - 0.4256 i} & {2.064e-6} & {0.6250 - 0.6622 i} & {1.773e-5} &  {- 0.1682 i} & {2.807e-11} \\
         10.5 & {0.3153 - 0.2005 i} & {1.644e-7} & {0.4682 - 0.4342 i} & {1.855e-6} & {0.6333 - 0.6749 i} & {1.591e-5} &  {- 0.1702 i} & {4.139e-11} \\
    \end{tabular}
    \end{ruledtabular}
\end{table*}

\begin{table*}
    \label{tab: AIM_q}
    \caption{QNM frequencies for the charge $q$. The results are calculated with asymptotic iteration method of $50$ iteration order.}
    \begin{ruledtabular}
    \begin{tabular}{ccccccccc}
         &\multicolumn{2}{c}{$\hat{\mathcal{n}}=0$}&\multicolumn{2}{c}{$\hat{\mathcal{n}}=1$}&\multicolumn{2}{c}{$\hat{\mathcal{n}}=2$}&\multicolumn{2}{c}{first pure imaginary}\\
         $q$&$\omega$&$\sqrt{\sigma_{re}^2+\sigma_{im}^2}$&$\omega$
        &$\sqrt{\sigma_{re}^2+\sigma_{im}^2}$&$\omega$
        &$\sqrt{\sigma_{re}^2+\sigma_{im}^2}$&$\omega$
        &$\sqrt{\sigma_{re}^2+\sigma_{im}^2}$\\ 
    \hline
         -0.5 & {0.2637 - 0.07208 i} & {2.256e-5} & {0.3336 - 0.1733 i} & {1.937e-4} & {0.4161 - 0.2830 i} & {2.627e-3} & { - 0.08566 i} & {7.629e-4} \\
         -1.5 & {0.2655 - 0.07258 i} & {1.282e-5} & {0.3385 - 0.1744 i} & {9.853e-5} & {0.4229 - 0.2854 i} & {8.608e-4} & { - 0.09169 i} & {3.496e-5} \\
         -2.5 & {0.2673 - 0.07315 i} & {4.219e-6} & {0.3432 - 0.1757 i} & {3.423e-5} & {0.4304 - 0.2886 i} & {3.392e-4} & { - 0.09728 i} & {1.314e-5} \\
         -3.5 & {0.2689 - 0.07377 i} & {5.536e-6} & {0.3478 - 0.1772 i} & {4.274e-5} & {0.4379 - 0.2908 i} & {3.733e-4} & { - 0.1022 i} & {1.449e-6} \\
         -4.5 & {0.2705 - 0.07444 i} & {3.660e-6} & {0.3522 - 0.1789 i} & {2.371e-5} & {0.4453 - 0.2931 i} & {1.997e-4} & { - 0.1067 i} & {1.089e-6} \\
         -5.5 & {0.2721 - 0.07513 i} & {2.591e-6} & {0.3562 - 0.1805 i} & {1.673e-5} & {0.4521 - 0.2960 i} & {1.263e-4} & { - 0.1107 i} & {5.847e-7} \\
         -6.5 & {0.2735 - 0.07584 i} & {2.131e-6} & {0.3601 - 0.1823 i} & {1.229e-5} & {0.4584 - 0.2987 i} & {8.281e-5} & { - 0.1143 i} & {4.157e-7} \\
         -7.5 & {0.2749 - 0.07656 i} & {1.768e-6} & {0.3638 - 0.1840 i} & {9.130e-6} & {0.4645 - 0.3015 i} & {5.755e-5} & { - 0.1177 i} & {3.843e-7} \\
         -8.5 & {0.2763 - 0.07728 i} & {3.659e-6} & {0.3673 - 0.1858 i} & {1.998e-5} & {0.4702 - 0.3043 i} & {1.293e-4} & { - 0.1208 i} & {1.040e-7} \\
         -9.5 & {0.2776 - 0.07801 i} & {3.190e-6} & {0.3707 - 0.1875 i} & {1.724e-5} & {0.4757 - 0.3069 i} & {1.081e-4} & { - 0.1237 i} & {7.619e-8} \\
         -10.5 & {0.2788 - 0.07874 i} & {2.887e-6} & {0.3739 - 0.1893 i} & {1.612e-5} & {0.4810 - 0.3097 i} & {9.016e-5} & { - 0.1264 i} & {5.360e-8} \\
         -11.5 & {0.2800 - 0.07946 i} & {1.136e-6} & {0.3770 - 0.1910 i} & {6.094e-6} & {0.4860 - 0.3125 i} & {2.472e-5} & { - 0.1290 i} & {1.778e-7} \\
         -12.5 & {0.2812 - 0.08018 i} & {1.069e-6} & {0.3800 - 0.1927 i} & {5.570e-6} & {0.4908 - 0.3152 i} & {2.389e-5} & { - 0.1314 i} & {1.560e-7} \\
         -13.5 & {0.2823 - 0.08089 i} & {1.019e-6} & {0.3828 - 0.1944 i} & {5.201e-6} & {0.4954 - 0.3178 i} & {2.295e-5} & { - 0.1337 i} & {1.291e-7} \\
         -14.5 & {0.2833 - 0.08159 i} & {2.287e-6} & {0.3856 - 0.1961 i} & {1.239e-5} & {0.4999 - 0.3205 i} & {5.623e-5} & { - 0.1359 i} & {4.780e-8} \\
         -15.5 & {0.2844 - 0.08229 i} & {2.215e-6} & {0.3882 - 0.1977 i} & {1.188e-5} & {0.5042 - 0.3231 i} & {5.508e-5} & { - 0.1380 i} & {4.064e-8} \\
         -16.5 & {0.2854 - 0.08298 i} & {2.164e-6} & {0.3908 - 0.1993 i} & {1.149e-5} & {0.5083 - 0.3256 i} & {5.465e-5} & { - 0.1400 i} & {3.685e-8} \\
         -17.5 & {0.2864 - 0.08366 i} & {2.122e-6} & {0.3933 - 0.2009 i} & {1.115e-5} & {0.5123 - 0.3281 i} & {5.450e-5} & { - 0.1419 i} & {3.757e-8} \\
         -18.5 & {0.2874 - 0.08433 i} & {2.081e-6} & {0.3957 - 0.2025 i} & {1.085e-5} & {0.5162 - 0.3306 i} & {5.445e-5} & { - 0.1437 i} & {4.074e-8} \\
         -19.5 & {0.2883 - 0.08500 i} & {2.040e-6} & {0.3981 - 0.2040 i} & {1.060e-5} & {0.5200 - 0.3330 i} & {5.442e-5} & { - 0.1455 i} & {4.393e-8} \\
         -20.5 & {0.2892 - 0.08565 i} & {1.997e-6} & {0.4003 - 0.2055 i} & {1.041e-5} & {0.5237 - 0.3354 i} & {5.437e-5} & { - 0.1472 i} & {4.591e-8} \\
    \end{tabular}
    \end{ruledtabular}
\end{table*}

\begin{table*}
    \caption{QNM frequencies for the Gauss-Bonnet coupling constant $\alpha$. The results are calculated with asymptotic iteration method of $50$ iteration order.}
    \begin{ruledtabular}
    \begin{tabular}{ccccccccc}
    \label{tab: AIM_alpha}
         &\multicolumn{2}{c}{$\hat{\mathcal{n}}=0$}&\multicolumn{2}{c}{$\hat{\mathcal{n}}=1$}&\multicolumn{2}{c}{$\hat{\mathcal{n}}=2$}&\multicolumn{2}{c}{first pure imaginary}\\
         $\alpha$&$\omega$&$\sqrt{\sigma_{re}^2+\sigma_{im}^2}$&$\omega$
        &$\sqrt{\sigma_{re}^2+\sigma_{im}^2}$&$\omega$
        &$\sqrt{\sigma_{re}^2+\sigma_{im}^2}$&$\omega$
        &$\sqrt{\sigma_{re}^2+\sigma_{im}^2}$\\ 
    \hline
         10 & {0.4567 - 0.1248 i} & {3.907e-5} & {0.5778 - 0.3001 i} & {3.356e-4} & {0.7207 - 0.4901 i} & {4.549e-3} &  {- 0.1484 i} & {1.321e-3} \\
         20 & {0.3230 - 0.08828 i} & {2.763e-5} & {0.4085 - 0.2122 i} & {2.373e-4} & {0.5096 - 0.3465 i} & {3.217e-3} &  {- 0.1049 i} & {9.343e-4} \\
         30 & {0.2637 - 0.07208 i} & {2.256e-5} & {0.3336 - 0.1733 i} & {1.937e-4} & {0.4161 - 0.2830 i} & {2.627e-3} &  {- 0.08566 i} & {7.629e-4} \\
         40 & {0.2284 - 0.06242 i} & {1.953e-5} & {0.2889 - 0.1501 i} & {1.678e-4} & {0.3604 - 0.2450 i} & {2.275e-3} &  {- 0.07419 i} & {6.607e-4} \\
         50 & {0.2043 - 0.05583 i} & {1.747e-5} & {0.2584 - 0.1342 i} & {1.501e-4} & {0.3223 - 0.2192 i} & {2.034e-3} &  {- 0.06635 i} & {5.909e-4} \\
         60 & {0.1865 - 0.05097 i} & {1.595e-5} & {0.2359 - 0.1225 i} & {1.370e-4} & {0.2942 - 0.2001 i} & {1.857e-3} &  {- 0.06057 i} & {5.394e-4} \\
         70 & {0.1726 - 0.04719 i} & {1.477e-5} & {0.2184 - 0.1134 i} & {1.268e-4} & {0.2724 - 0.1852 i} & {1.719e-3} &  {- 0.05608 i} & {4.994e-4} \\
         80 & {0.1615 - 0.04414 i} & {1.381e-5} & {0.2043 - 0.1061 i} & {1.186e-4} & {0.2548 - 0.1733 i} & {1.608e-3} &  {- 0.05246 i} & {4.672e-4} \\
         90 & {0.1522 - 0.04161 i} & {1.302e-5} & {0.1926 - 0.1000 i} & {1.119e-4} & {0.2402 - 0.1634 i} & {1.516e-3} &  {- 0.04946 i} & {4.405e-4} \\
         100 & {0.1444 - 0.03948 i} & {1.235e-5} & {0.1827 - 0.09491 i} & {1.061e-4} & {0.2279 - 0.1550 i} & {1.439e-3} &  {- 0.04692 i} & {4.179e-4} \\
    \end{tabular}
    \end{ruledtabular}
\end{table*}

\begin{table*}
    \label{tab: AIM_l}
    \caption{QNM frequencies for the ``quantum number" $l$. The results are calculated with asymptotic iteration method of $50$ iteration order. "-" symbol indicates that AIM can't predict the result with enough precision with corresponding parameter choice.}
    \begin{ruledtabular}
    \begin{tabular}{ccccccccc}
         &\multicolumn{2}{c}{$\hat{\mathcal{n}}=0$}&\multicolumn{2}{c}{$\hat{\mathcal{n}}=1$}&\multicolumn{2}{c}{$\hat{\mathcal{n}}=2$}&\multicolumn{2}{c}{first pure imaginary}\\
         $l$&$\omega$&$\sqrt{\sigma_{re}^2+\sigma_{im}^2}$&$\omega$
        &$\sqrt{\sigma_{re}^2+\sigma_{im}^2}$&$\omega$
        &$\sqrt{\sigma_{re}^2+\sigma_{im}^2}$&$\omega$
        &$\sqrt{\sigma_{re}^2+\sigma_{im}^2}$\\ 
    \hline
         0 & {0.1294 - 0.1119 i} & {1.036e-3} & - & - & - & - &  {- 0.03624 i} & {1.374e-6} \\
         1 & {0.1389 - 0.1065 i} & {1.021e-3} & - & - & - & - &  {- 0.04079 i} & {3.095e-6} \\
         2 & {0.1558 - 0.09809 i} & {4.608e-4} & {0.2553 - 0.2135 i} & {5.634e-3} & - & - &  {- 0.04817 i} & {1.112e-6} \\
         3 & {0.1790 - 0.08950 i} & {2.309e-4} & {0.2644 - 0.2023 i} & {2.541e-3} & - & - &  {- 0.05678 i} & {1.939e-6} \\
         4 & {0.2057 - 0.08262 i} & {4.372e-5} & {0.2859 - 0.1909 i} & {4.925e-4} & - & - &  {- 0.06592 i} & {6.728e-5} \\
         5 & {0.2342 - 0.07686 i} & {4.796e-5} & {0.3085 - 0.1816 i} & {4.646e-4} & {0.3923 - 0.2903 i} & {5.414e-3} &  {- 0.07547 i} & {1.062e-4} \\
         6 & {0.2637 - 0.07208 i} & {2.256e-5} & {0.3336 - 0.1733 i} & {1.937e-4} & {0.4161 - 0.2830 i} & {2.627e-3} &  {- 0.08566 i} & {7.629e-4} \\
         7 & {0.2938 - 0.06795 i} & {1.061e-5} & {0.3598 - 0.1654 i} & {8.851e-5} & {0.4374 - 0.2761 i} & {1.737e-3} & - & - \\
         8 & {0.3243 - 0.06437 i} & {5.454e-6} & {0.3873 - 0.1583 i} & {4.575e-5} & {0.4603 - 0.2660 i} & {1.129e-3} & - & - \\
         9 & {0.3551 - 0.06123 i} & {3.162e-6} & {0.4156 - 0.1519 i} & {2.521e-5} & {0.4858 - 0.2559 i} & {6.110e-4} & - & - \\
         10 & {0.3860 - 0.05844 i} & {2.322e-6} & {0.4446 - 0.1461 i} & {2.232e-5} & {0.5126 - 0.2475 i} & {5.824e-4} & - & - \\
         11 & {0.4172 - 0.05594 i} & {1.747e-6} & {0.4740 - 0.1408 i} & {1.720e-5} & {0.5402 - 0.2398 i} & {6.290e-4} & - & - \\
         12 & {0.4484 - 0.05369 i} & {1.255e-6} & {0.5037 - 0.1361 i} & {1.849e-5} & {0.5679 - 0.2324 i} & {7.605e-4} & - & - \\
         13 & {0.4798 - 0.05163 i} & {8.882e-7} & {0.5337 - 0.1317 i} & {2.403e-5} & {0.5953 - 0.2257 i} & {1.029e-3} & - & - \\
         14 & {0.5112 - 0.04974 i} & {6.810e-7} & {0.5640 - 0.1277 i} & {3.519e-5} & {0.6238 - 0.2211 i} & {1.423e-3} & - & - \\
         15 & {0.5427 - 0.04800 i} & {7.710e-7} & {0.5944 - 0.1240 i} & {5.292e-5} & {0.6554 - 0.2161 i} & {1.990e-3} & - & - \\
         16 & {0.5743 - 0.04639 i} & {1.209e-6} & {0.6250 - 0.1204 i} & {8.011e-5} & {0.6867 - 0.2077 i} & {2.809e-3} & - & - \\
         17 & {0.6059 - 0.04490 i} & {1.995e-6} & {0.6555 - 0.1173 i} & {1.213e-4} & {0.7132 - 0.1991 i} & {3.881e-3} & - & - \\
         18 & {0.6375 - 0.04349 i} & {9.371e-7} & {0.6864 - 0.1143 i} & {5.338e-5} & {0.7420 - 0.1972 i} & {1.647e-3} & - & - \\
         19 & {0.6692 - 0.04218 i} & {1.541e-6} & {0.7174 - 0.1116 i} & {8.297e-5} & {0.7700 - 0.1947 i} & {2.385e-3} & - & - \\
         20 & {0.7009 - 0.04094 i} & {2.534e-6} & {0.7485 - 0.1088 i} & {1.279e-4} & {0.8011 - 0.1956 i} & {3.615e-3} & - & - \\
    \end{tabular}
    \end{ruledtabular}
\end{table*}

\begin{table*}
    \caption{QNM frequencies for the ``quantum number" $\gamma$. The results are calculated with asymptotic iteration method of $50$ iteration order.}
    \begin{ruledtabular}
    \begin{tabular}{ccccccccc}
    	\label{tab: AIM_gamma}
         &\multicolumn{2}{c}{$\hat{\mathcal{n}}=0$}&\multicolumn{2}{c}{$\hat{\mathcal{n}}=1$}&\multicolumn{2}{c}{$\hat{\mathcal{n}}=2$}&\multicolumn{2}{c}{first pure imaginary}\\
         $\gamma$&$\omega$&$\sqrt{\sigma_{re}^2+\sigma_{im}^2}$&$\omega$
        &$\sqrt{\sigma_{re}^2+\sigma_{im}^2}$&$\omega$
        &$\sqrt{\sigma_{re}^2+\sigma_{im}^2}$&$\omega$
        &$\sqrt{\sigma_{re}^2+\sigma_{im}^2}$\\ 
    \hline
         0 & {0.2637 - 0.07207 i} & {2.841e-5} & {0.3336 - 0.1733 i} & {2.460e-4} & {0.4155 - 0.2821 i} & {2.514e-3} &  {- 0.08552 i} & {5.121e-4} \\
         1 & {0.2622 - 0.06978 i} & {3.295e-5} & {0.3317 - 0.1703 i} & {2.749e-4} & {0.4137 - 0.2790 i} & {2.685e-3} &  {- 0.08446 i} & {5.226e-4} \\
         2 & {0.2606 - 0.06742 i} & {3.875e-5} & {0.3298 - 0.1672 i} & {3.117e-4} & {0.4116 - 0.2758 i} & {2.862e-3} &  {- 0.08342 i} & {5.333e-4} \\
         3 & {0.2590 - 0.06500 i} & {4.122e-5} & {0.3278 - 0.1640 i} & {3.203e-4} & {0.4092 - 0.2732 i} & {2.844e-3} &  {- 0.08250 i} & {6.506e-4} \\
         4 & {0.2573 - 0.06248 i} & {4.470e-5} & {0.3256 - 0.1606 i} & {3.343e-4} & {0.4059 - 0.2704 i} & {2.878e-3} &  {- 0.08162 i} & {7.885e-4} \\
         5 & {0.2556 - 0.05985 i} & {5.582e-5} & {0.3233 - 0.1571 i} & {3.998e-4} & {0.4027 - 0.2661 i} & {3.097e-3} &  {- 0.08064 i} & {8.070e-4} \\
         6 & {0.2537 - 0.05708 i} & {5.686e-5} & {0.3207 - 0.1533 i} & {3.817e-4} & {0.3984 - 0.2616 i} & {3.241e-3} &  {- 0.07988 i} & {1.203e-3} \\
         7 & {0.2517 - 0.05415 i} & {6.728e-5} & {0.3180 - 0.1492 i} & {4.216e-4} & {0.3959 - 0.2566 i} & {3.573e-3} &  {- 0.07887 i} & {1.475e-3} \\
         8 & {0.2496 - 0.05102 i} & {9.341e-5} & {0.3150 - 0.1448 i} & {5.564e-4} & {0.3940 - 0.2521 i} & {3.907e-3} &  {- 0.07771 i} & {1.437e-3} \\
         9 & {0.2472 - 0.04759 i} & {1.216e-4} & {0.3116 - 0.1401 i} & {6.794e-4} & {0.3917 - 0.2482 i} & {4.259e-3} &  {- 0.07640 i} & {1.634e-3} \\
         10 & {0.2446 - 0.04381 i} & {2.204e-4} & {0.3077 - 0.1347 i} & {1.172e-3} & {0.3898 - 0.2448 i} & {6.182e-3} &  {- 0.07529 i} & {1.350e-3} \\
         11 & {0.2416 - 0.03930 i} & {4.068e-4} & {0.3025 - 0.1285 i} & {2.016e-3} & {0.3882 - 0.2437 i} & {9.990e-3} &  {- 0.07406 i} & {1.348e-3} \\
    \end{tabular}
    \end{ruledtabular}
\end{table*}

\section{conclusions and discussion}\label{sec: conclusions_and_discussion}
In this paper, we investigate the QNMs of tensor perturbation for the Kaluza-Klein black hole in EGB gravity. The topology of the so-called Kaluza-Klein black hole is the product of a usual $4$-dimensional spacetime with a negative constant curvature space. The metric function $f(r)$ of the black hole with $k=1$ is determined by four parameters. These parameters are the dimension of spacetime $n$, the Gauss-Bonnet coupling constant $\alpha$, the mass parameter $\mu$ and the charge parameter $q$, respectively. From the asymptotic expansion of the metric function, it is found that the behavior of the metric function is similar to the Reissner-Nordstrom-anti-de Sitter spacetime while the Maxwell field is absent.

The establishment of the tensor perturbation for the Kaluza-Klein black hole in EGB gravity is tedious. In our previous work, we have got the Kodama-Ishibashi formalism for the tensor perturbation of the theory, and a generalized master equations are given~\cite{Cao:2021sty}. The applicability of this master equation is broad. Indeed, it can be used to calculate the perturbation of all tensor types for warped product spacetimes in EGB gravity theory. Evidently, it is applicable to the spacetime studied in this paper. It should be emphasized that our perturbation equation or the Schr\"{o}dinger-like equation is based on the equations of motion of EGB theory and the corrections of the coefficients of the second order covariant derivative, i.e., the term $\sim\alpha G^{ab}$, arises naturally. Different from a simple test field equation, this perturbation equation can reveal the dynamics of underlying gravitational theories beyond the given geometry of spacetime. To illustrate this difference, we give a toy model, i.e., Klein-Gordon test field on the four dimensional spacetime in Appendix \ref{KG}. The difference in the result verify the influence of the dynamics on the compactification part on the 4-dimensional spacetime. Once detected, this might serve as a probe for the existence of extra dimension.

Using the asymptotic iteration method (AIM) and the numerical integration method, we calculated the QNMs for different parameters. As for the AIM, considering that there is limited literature on how to select expansion points, we propose a method for selecting the expansion point $\xi_0$ in Sec. \ref{sec: AIM}. The first three order characteristic frequencies obtained from AIM including the first pure imaginary modes are shown in the Sec. \ref{sec: numerical_result_and_analysis}. The corresponding monotonicities are also described for the parameters $n$, $\mu$, $q$, $\alpha$, $l$ and $\gamma$. Furthermore, to demonstrate the accuracies of characteristic frequencies, we provide the formula of the relative error between the results of the two methods. The conclusion is that the errors are relatively small.

Technically speaking, due to limitations in computational accuracy, it is difficult for us to find the QNM frequencies with $\hat{\mathcal{n}}\ge 3$. The continued fractions method or Leaver method is a great benefit for getting more overtones of QNMs~\cite{Leaver:1985ax,Leaver:1990zz,Nollert:1993zz,Konoplya:2004wg,Zhidenko:2006rs,Daghigh:2022uws}. However, the metric function $f(r)$ is not a rational function so that if one use this method, an infinite recurrence relation will be obtained. Fortunately, Rezzolla and Zhidenko use continue fractions expansions in terms of the metric function to overcome it~\cite{Rezzolla:2014mua}. Recently, Konoplya {\it et al.} developed a general procedure allowing one to use the Leaver method to metrics which are not expressed in terms of rational functions~\cite{Konoplya:2023aph}. Inspired by this work, we aim to do further calculations with this delicate method to get more overtones in future work.

The undetermined parameters of the theory can be constrained by comparing the results of our calculations and predictions with data from actual observations. However, there are several undetermined parameters in the theory, including total spacetime dimensions $n$, Gauss-Bonnet coupling parameter $\alpha$, and eigenvalues $\gamma$ of the characteristic tensors. This complexity makes their determination quite challenging. Here, we present a potential method for imposing constraints on these parameters. The product $\sqrt{\alpha}\omega$ is dimensionless and is independent of $\alpha$ and can be inferred from the results we have computed. Therefore we can rewrite it as
    \begin{eqnarray}
        \sqrt{\alpha}\omega=F(n, \mu, q, l, \gamma)\, .
    \end{eqnarray}
    Consider the dimensionless quantity,
    \begin{eqnarray}
    M_{\text{eff}}\,\omega&=&\dfrac{1}{2}\alpha^{1/2}\omega\mu\sqrt{3(n-4)(n-5)}\nonumber\\
    &=&\dfrac{1}{2}F(n, \mu, q, l, \gamma)\mu\sqrt{3(n-4)(n-5)}\, ,
    \end{eqnarray}
    where the absence of any dependence on $\alpha$ should be noted. One can compare this quantity with the actual observed quantity and impose constraint on $n, \mu, q, l, \gamma$, and therefore determine the potential specific value of $F$, i.e., $\sqrt{\alpha}\omega$. Once the precise value of $\sqrt{\alpha}\omega$ is determined, one can leverage the observed value of $\omega$ to obtain the specific value of $\alpha$. However, to do this specifically requires a lot of effort and is beyond the scope of our present work. Nevertheless, we may limit these parameters through the methods given above in the later studies.
    
Another point worth noting is that by the use of the gauge-invariant variables proposed by Kodama and Ishibashi~\cite{Ishibashi:2011ws}, the most general perturbation equations of General Relativity in the $(m+n)$-dimensional spacetime with a warped product metric has been obtained in~\cite{Cai:2013cja}. In the EGB gravity theory, it is worthwhile to use similar methods to get the master equations of the scalar and vector type for the $(m+n)$-dimensional spacetime. With these equations, we can calculate the QNMs under different black hole backgrounds. In the frame of EGB gravity theory with $(m+n)$-dimensional spacetime, computing the perturbation equations and QNMs are both non-trivial things which will be considered in the future.

\section*{Acknowledgement}
We are grateful to Yi-Fu Cai for helpful discussions. This work was supported in part by the National Natural Science Foundation of China with grants No.12075232, No.12247103 and No.11961131007. This work is also supported by the Fundamental Research Funds for the Central Universities under Grant No.WK2030000036, and Grant NO.WK2030000044. This work is supported by the National Key R\&D Program of China Grant No. 2022YFC2204603. Part of numerics were operated on the computer clusters {\it LINDA} \& {\it JUDY} in the particle cosmology group at USTC.

\appendix
\section{The four dimensional scalar equation}
\label{app_1}
In this Appendix, we will show how the four dimensional scalar equation (\ref{master_equation_4d}) is derived from the master equation of the tensor perturbation (\ref{master_equation}). Substituting Eq. (\ref{separate_variables}) into Eq. (\ref{master_equation}), we have
\begin{eqnarray}
	(P^{ab}D_aD_b+P^{mn}\hat{D}_m\hat{D}_n+P^aD_a+V)\Big[\Phi(y)\bar{h}_{ij}\Big]=0\, .
\end{eqnarray}
With the help of Eq. (\ref{gamma}), a scalar equation in the four dimensional manifold $M^4$
\begin{eqnarray}\label{master_equation_2}
	\Big(P^{ab}D_aD_b+P^aD_a+V+\frac{Q\gamma}{r^2}\Big)\Phi(y)=0\, ,
\end{eqnarray}
is obtained. Note that $r=r_0$, the terms with derivatives with respect to $r$ are all vanished. Therefore, from Eqs. (\ref{Pab})-(\ref{V}), we have
\begin{gather}
	P^{ab}=\frac{4n-22}{(n-4)(n-5)}g^{ab}-4\alpha\cdot{}^4\! G^{ab}\, ,
    \\
	Q=\frac{6(n-6)}{(n-4)(n-5)}+2\alpha\cdot{}^4\!{R}\, ,
    \\
	P^{a}=0\, ,
\end{gather}
and
\begin{eqnarray}
	V=\frac{2}{(n-4)(n-5)}{}^4\!{R}+\frac{6(n-6)}{\alpha(n-4)^2(n-5)^2}\, ,
\end{eqnarray}
where we have used Eq. (\ref{r0}) and Eq. (\ref{KK_main_equation}) with a relation between $\alpha$ and $\Lambda$, i.e.,
\begin{eqnarray}\label{Lambda}
	\alpha\Lambda=-\frac{n^2-5n-2}{8(n-4)(n-5)}\, .
\end{eqnarray}
It should be noted that since $r_0^2>0$ and $\alpha>0$,  we have $K=-1$, and $\Lambda<0$. Hence, Eq. (\ref{master_equation_2}) becomes
\begin{eqnarray}
	\Big[\frac{4n-22}{(n-4)(n-5)}g^{ab}-4\alpha\cdot{}^4\! G^{ab}\Big]D_aD_b\Phi+\Big[\frac{2+\gamma}{(n-4)(n-5)}{}^4\!{R}+\frac{3(n-6)(2+\gamma)}{\alpha(n-4)^2(n-5)^2}\Big]\Phi=0\, .
\end{eqnarray}

\section{The proof of $V_{\text{eff}}(r_{+})=0$}\label{app_2}
In this Appendix, we will give a proof of $V_{\text{eff}}(r_{+})=0$. First, we have the function $B(r)$ given by Eq. (\ref{B(r)}). As $r\to r_{+}$, the function $f\to0$. On the one hand,
\begin{eqnarray}\label{proof_1}
	&&\lim_{r\to r_{+}}\frac{f^2(r)B^2(r)}{4}\nonumber\\
	&=&\lim_{r\to r_{+}}\frac{f^2}{4}\Big[\frac{4n-22}{(n-4)(n-5)}\Big(f^{\prime}+\frac{2f}{r}\Big)-4\alpha\frac{-f^{\prime}+3ff^{\prime}+r(f^{\prime})^2+rff^{\prime\prime}}{r^2}\Big]^2\times\nonumber\\
	&&\Big[\frac{4n-22}{(n-4)(n-5)}f-\frac{4\alpha f(-1+f+rf^{\prime})}{r^2}\Big]^{-2}\nonumber\\
	&=&\frac14\lim_{r\to r_{+}}\Big[\frac{4n-22}{(n-4)(n-5)}f^{\prime}-4\alpha\frac{-f^{\prime}+r(f^{\prime})^2}{r^2}\Big]^2\Big[\frac{4n-22}{(n-4)(n-5)}-\frac{4\alpha(-1+rf^{\prime})}{r^2}\Big]^{-2}\nonumber\\
	&=&\frac{(f^{\prime}(r_{+}))^2}{4}\, .
\end{eqnarray}
On the other hand, the derivative of $B(r)$ is 
\begin{eqnarray}
	B^{\prime}(r)&=&\frac{1}{f^2\Big[\frac{4 n-22}{(n-5)(n-4)}-\frac{4\alpha(r f^{\prime}+f-1)}{r^2}\Big]^2}\Biggl\{-\Big[\frac{(4 n-22) f^{\prime}}{(n-5)(n-4)}+\frac{8 \alpha f(r f^{\prime}+f-1)}{r^3}-\frac{4 \alpha (r f^{\prime}+f-1) f^{\prime}}{r^2}\nonumber\\
	&&-\frac{4\alpha f(rf^{\prime\prime}+2 f^{\prime})}{r^2}\Big]\times\Big[\frac{(4 n-22)\Big(f^{\prime}+\frac{2 f}{r}\Big)}{(n-5) (n-4)}-\frac{4 \alpha (r ff^{\prime\prime}+r (f^{\prime})^2+(3f-1)f^{\prime})}{r^2}\Big]\nonumber\\
	&&+f\Big[\frac{4 n-22}{(n-5)(n-4)}-\frac{4 \alpha  (r f^{\prime}+f-1)}{r^2}\Big]\times\Big[\frac{(4n-22)\Big(f^{\prime\prime}+\frac{2 f^{\prime}}{r}-\frac{2 f}{r^2}\Big)}{(n-5)(n-4)}+\frac{8 \alpha (r ff^{\prime\prime}+r(f^{\prime})^2+(3 f-1) f^{\prime})}{r^3}\nonumber\\
	&&-\frac{4\alpha(rff^{(3)}+(4 f-1)f^{\prime\prime}+4(f^{\prime})^2+3 r f^{\prime}f^{\prime\prime})}{r^2}\Big]\Biggr\}\, .
\end{eqnarray}
Therefore, we have
\begin{eqnarray}\label{proof_2}
	&&\lim_{r\to r_{+}}\frac{f^2(r)B^{\prime}(r)}{2}\nonumber\\
	&=&\lim_{r\to r_{+}}\frac12\Big[\frac{4 n-22}{(n-5)(n-4)}-\frac{4\alpha(r f^{\prime}+f-1)}{r^2}\Big]^{-2}\Biggl\{-\Big[\frac{(4 n-22) f^{\prime}}{(n-5)(n-4)}+\frac{8 \alpha f(r f^{\prime}+f-1)}{r^3}-\frac{4 \alpha (r f^{\prime}+f-1) f^{\prime}}{r^2}\nonumber\\
	&&-\frac{4\alpha f(rf^{\prime\prime}+2 f^{\prime})}{r^2}\Big]\times\Big[\frac{(4 n-22)\Big(f^{\prime}+\frac{2 f}{r}\Big)}{(n-5) (n-4)}-\frac{4 \alpha (r ff^{\prime\prime}+r (f^{\prime})^2+(3f-1)f^{\prime})}{r^2}\Big]\nonumber\\
	&&+f\Big[\frac{4 n-22}{(n-5)(n-4)}-\frac{4 \alpha  (r f^{\prime}+f-1)}{r^2}\Big]\times\Big[\frac{(4n-22)\Big(f^{\prime\prime}+\frac{2 f^{\prime}}{r}-\frac{2 f}{r^2}\Big)}{(n-5)(n-4)}+\frac{8 \alpha (r ff^{\prime\prime}+r(f^{\prime})^2+(3 f-1) f^{\prime})}{r^3}\nonumber\\
	&&-\frac{4\alpha(rff^{(3)}+(4 f-1)f^{\prime\prime}+4(f^{\prime})^2+3 r f^{\prime}f^{\prime\prime})}{r^2}\Big]\Biggr\}\nonumber\\
	&=&\lim_{r\to r_{+}}\frac12\Big[\frac{4 n-22}{(n-5)(n-4)}-\frac{4\alpha(r f^{\prime}-1)}{r^2}\Big]^{-2}\Biggl\{-\Big[\frac{(4 n-22) f^{\prime}}{(n-5)(n-4)}-\frac{4 \alpha (r f^{\prime}-1) f^{\prime}}{r^2}\Big]\Big[\frac{(4 n-22)f^{\prime}}{(n-5) (n-4)}-\frac{4 \alpha (r (f^{\prime})^2-f^{\prime})}{r^2}\Big]\Biggr\}\nonumber\\
	&=&-\frac{(f^{\prime}(r_{+}))^2}{2}\, .
\end{eqnarray}
From Eq. (\ref{proof_1}) and Eq. (\ref{proof_2}), as $r\to r_{+}$, the limit of $V_{\text{eff}}$  is
\begin{eqnarray}
	V_{\text{eff}}(r_{+})&=&\lim_{r\to r_{+}}\Big[\omega^2-f^2C+\frac{(f^{\prime})^2}{4}-\frac{ff^{\prime\prime}}{2}+\frac{f^2B^{\prime}}{2}+\frac{f^2B^2}{4}\Big]\nonumber\\
	&=&\frac{(f^{\prime}(r_{+}))^2}{4}-\frac{(f^{\prime}(r_{+}))^2}{2}+\frac{(f^{\prime}(r_{+}))^2}{4}=0\, .
\end{eqnarray}

\section{The asymptotic behavior of $\varphi$}\label{app_3}
In this Appendix, we will look at the asymptotic behavior of $\varphi$ in order to apply the boundary condition (\ref{boundary_condition}). In terms of $r$, consider that $\mathrm{d}r_{\star}=\mathrm{d}r/f$, Eq. (\ref{Schrodinger_equation}) becomes
\begin{eqnarray}\label{Schrodinger_2}
	\varphi^{\prime\prime}+p(r)\varphi^{\prime}+q(r)\varphi=0\, ,
\end{eqnarray}
where
\begin{eqnarray}
	p(r)=\frac{f^{\prime}(r)}{f(r)}\, ,\quad q(r)=\frac{\omega^2-V_{\text{eff}}(r)}{f^2(r)}\, .
\end{eqnarray}
For the sake of finding out the the asymptotic behavior of $\varphi$, there is a useful theorem~\cite{wang1989special}:

{\it The necessary and sufficient condition for Eq. (\ref{Schrodinger_2}) to have two regular solutions in the neighborhood $0<|r-r_0|<\delta$ of its singular point $r_0$ is that the functions 
\begin{eqnarray}
	(r-r_0)p(r)\, ,\quad (r-r_0)^2q(r)
\end{eqnarray}
are both are all analytic in $|r-r_{+}|<\delta$.
}

The singular point satisfied with the theorem is called regular singular point. Otherwise, the points are called irregular singular point. 
Since we are considering the non-degenerate case, i.e., $f^{\prime}(r_{+})\neq0$, so it is easy to find that
\begin{eqnarray}
	(r-r_{+})p(r)=(r-r_{+})\frac{f^{\prime}(r)}{f(r)}\quad \text{and}\quad (r-r_{+})^2q(r)=(r-r_{+})^2\frac{\omega^2-V_{\text{eff}}(r)}{f^2(r)}
\end{eqnarray}
are both analytic in $|r-r_{+}|<\delta$. Hence, $r=r_{+}$ is the regular singular point. The index equation is given by
\begin{eqnarray}\label{index_equation}
	\rho(\rho-1)+a_0\rho+b_0=0\, ,
\end{eqnarray}
where
\begin{eqnarray}
	a_0=\lim_{r\to r_{+}}(r-r_{+})p(r)=\lim_{r\to r_{+}}(r-r_{+})\frac{f^{\prime}(r)}{f(r)}=1\, ,
\end{eqnarray}
\begin{eqnarray}
	b_0=\lim_{r\to r_{+}}(r-r_{+})^2q(r)=\lim_{r\to r_{+}}(r-r_{+})^2\frac{\omega^2-V_{\text{eff}}(r)}{f^2(r)}=\frac{\omega^2}{(f^{\prime}(r_{+}))^2}\, .
\end{eqnarray}
The index equation (\ref{index_equation}) becomes
\begin{eqnarray}
	\rho^2+\frac{\omega^2}{(f^{\prime}(r_{+}))^2}=0\, .
\end{eqnarray}
Therefore, we have $\rho=\pm i\omega/f^{\prime}(r_{+})$. Considering the boundary condition, we have the asymptotic behavior of $\varphi$ at $r\to r_{+}$ reading as
\begin{eqnarray}
	\varphi\sim\Big(\frac{r-r_{+}}{r-r_{-}}\Big)^{-i\omega/f^{\prime}(r_{+})}\, .
\end{eqnarray}
As for the behavior of $r\to+\infty$, define $t=1/r$, and then Eq. (\ref{Schrodinger_2}) becomes
\begin{eqnarray}\label{Schrodinger_3}
	\frac{\mathrm{d}^2\varphi}{\mathrm{d}t^2}+\Big[\frac{2}{t}-\frac{1}{t^2}p\Big(\frac{1}{t}\Big)\Big]\frac{\mathrm{d}\varphi}{\mathrm{d}t}+\frac{1}{t^4}q\Big(\frac{1}{t}\Big)\varphi(t)=0\, .
\end{eqnarray}
Since we have
\begin{eqnarray}
	\lim_{t\to0}\frac{1}{t}p\Big(\frac{1}{t}\Big)=\lim_{r\to+\infty}\frac{rf^{\prime}(r)}{f(r)}=\lim_{r\to+\infty}\frac{r\cdot\frac{2r}{2(n-4)\alpha}\Big[1-\sqrt{\frac{n-4}{3(n-5)}}\Big]}{\frac{r^2}{2(n-4)\alpha}\Big[1-\sqrt{\frac{n-4}{3(n-5)}}\Big]}=2\, ,
\end{eqnarray}
and
\begin{eqnarray}
	\lim_{t\to0}\frac{1}{t^2}q\Big(\frac{1}{t}\Big)=\lim_{r\to+\infty}\frac{r^2(\omega^2-V_{\text{eff}}(r))}{f^2(r)}=\lim_{r\to+\infty}\frac{-r^2\cdot V_0r^2}{\Big\{\frac{r^2}{2(n-4)\alpha}\Big[1-\sqrt{\frac{n-4}{3(n-5)}}\Big]\Big\}^2}=-\frac{4(n-4)^2\alpha^2V_0}{\Big[1-\sqrt{\frac{n-4}{3(n-5)}}\Big]^2}\, ,
\end{eqnarray}
as a result, $r=\infty$ is a regular singular point. Now, $a_0$ and $b_0$ in the index equation (\ref{index_equation}) are given by
\begin{eqnarray}
	a_0=\lim_{t\to0}t\Big[\frac{2}{t}-\frac{1}{t^2}p\Big(\frac{1}{t}\Big)\Big]=2-\lim_{t\to0}\frac{1}{t}p\Big(\frac{1}{t}\Big)=0\, ,
\end{eqnarray}
and
\begin{eqnarray}
	b_0=\lim_{t\to0}t^2\Big[\frac{1}{t^4}q\Big(\frac{1}{t}\Big)\Big]=\lim_{t\to0}\frac{1}{t^2}q\Big(\frac{1}{t}\Big)=-\frac{4(n-4)^2\alpha^2V_0}{\Big[1-\sqrt{\frac{n-4}{3(n-5)}}\Big]^2}\, .
\end{eqnarray}
The condition $V_0>0$ is required in our paper. So, we obtain $b_0<0$. There are two different roots of the index equation $\rho^2-\rho+b_0=0$ with $b_0<0$. The roots are
\begin{eqnarray}
	\rho_1=\frac{1+\sqrt{1-4b_0}}{2}>0\, ,
\end{eqnarray}
and
\begin{eqnarray}
	\rho_2=\frac{1-\sqrt{1-4b_0}}{2}<0\, .
\end{eqnarray}
The boundary condition is $\varphi(t)\to0$ as $t\to0$. Therefore, $\rho=\rho_1>0$ is requisite. For convenience, we define
\begin{eqnarray}
	\rho\equiv\rho_1=\frac{1}{2}\Biggl\{1+\sqrt{1+\frac{16(n-4)^2\alpha^2V_0}{\Big[1-\sqrt{\frac{n-4}{3(n-5)}}\Big]^2}}\Biggr\}\, .
\end{eqnarray}
In order to apply the boundary condition (\ref{boundary_condition}), we define the following solution
\begin{eqnarray}
	\varphi(r)=\Big(\frac{r-r_{+}}{r-r_{-}}\Big)^{-i\omega/f^{\prime}(r_{+})}\Big(\frac{r_{+}-r_{-}}{r-r_{-}}\Big)^{\rho}\tilde{\varphi}(r)\, .
\end{eqnarray}

\section{The toy model: the Klein-Gordon equation}\label{KG}
\begin{figure}[htbp]
	\centering
	\includegraphics[width=0.33\textwidth]{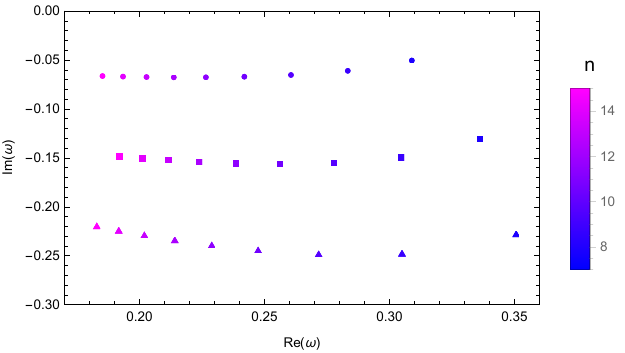}
    \includegraphics[width=0.33\textwidth]{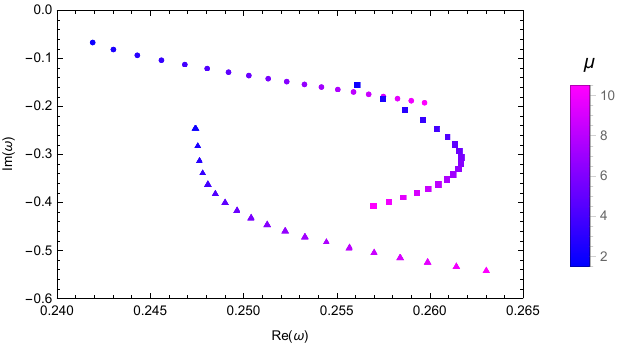}
    \includegraphics[width=0.33\textwidth]{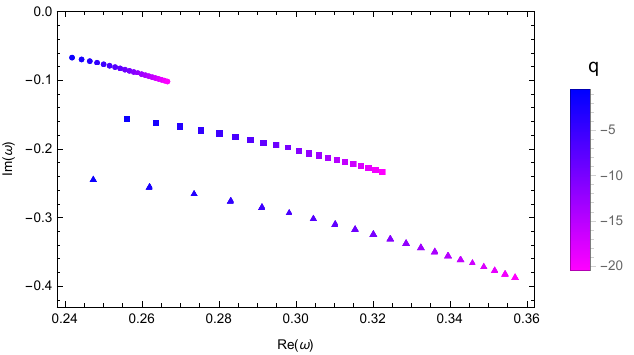}
    \includegraphics[width=0.33\textwidth]{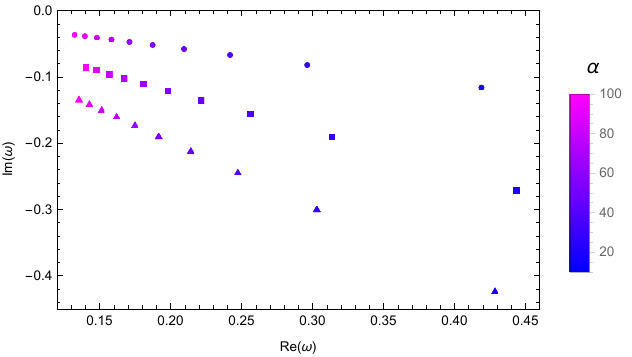}
    \includegraphics[width=0.33\textwidth]{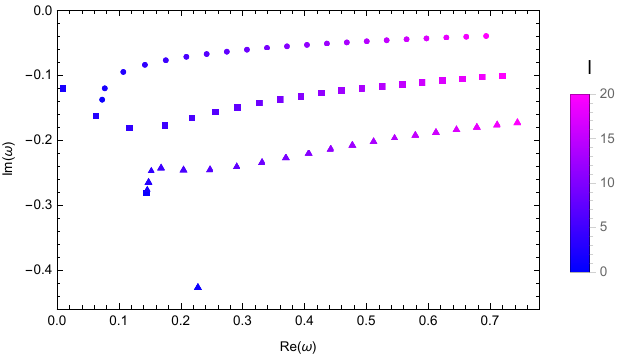}
	\caption{QNM frequencies of overtone number $\hat{\mathcal{n}}=0, 1, 2$ for the Klein-Gordon equation with different physical parameters. The six figures are for the spacetime dimension $n$, the mass $\mu$, the charge $q$, the Gauss-Bonnet coupling constant $\alpha$, and the ``quantum numbers" $l$ and $\gamma$. Results with overtone number $\hat{\mathcal{n}}=0, 1, 2$ are marked by circle, rectangle and triangle, repesctively. }
	\label{fig: AIM_KG_data_graph}
\end{figure}

In this Appendix, we use the Klein-Gordon equation as a toy model to compute the QNM frequencies in the four dimensional spacetime $(M^4,g_{ab})$. In this case, the effective potential has a simple form
\begin{eqnarray}
	V_{\text{eff}}(r)=f(r)\Big[\frac{l(l+1)}{r^2}+\frac{f^{\prime}(r)}{r}\Big]\, ,
\end{eqnarray}
where the metric is given by Eq. (\ref{f(r)_k_1}). After considering the boundary conditions of QNMs, we define the following solution
\begin{eqnarray}
	\varphi(r)=\Big(\frac{r-r_{+}}{r-r_{-}}\Big)^{-i\omega/f^{\prime}(r_{+})}\Big(\frac{r_{+}-r_{-}}{r-r_{-}}\Big)^{2}\tilde{\varphi}(r)\, .
\end{eqnarray}
Substituting the above expression into the AIM algorithm, we get the QNM frequencies shown in Fig. \ref{fig: AIM_KG_data_graph}.

\bibliography{reference}{}
\bibliographystyle{apsrev4-1}

\end{document}